\documentclass[pre,tighten]{revtex4}
\usepackage[cp850]{inputenc}
\usepackage{natbib}
\usepackage{amssymb,amsmath}
\usepackage{epsfig}
\usepackage{bm}
\usepackage{color}

\usepackage{graphicx}% Include figure files

\usepackage{color}
\newcommand{\vicente}[1]{{ #1}}
\newcommand\beq{\begin{equation}}
\newcommand\eeq{\end{equation}}
\newcommand\beqa{\begin{eqnarray}}
\newcommand\eeqa{\end{eqnarray}}

\newcommand{\al}{\alpha}

\begin{document}

\title{Transport coefficients for granular suspensions at moderate densities}

\author{Rub\'en G\'omez Gonz\'alez\footnote[1]{Electronic address: ruben@unex.es}}
\affiliation{Departamento de F\'{\i}sica,
Universidad de Extremadura, E-06006 Badajoz, Spain}
\author{Vicente Garz\'{o}\footnote[2]{Electronic address: vicenteg@unex.es;
URL: http://www.unex.es/eweb/fisteor/vicente/}}
\affiliation{Departamento de F\'{\i}sica and Instituto de Computaci\'on Cient\'{\i}fica Avanzada (ICCAEx), Universidad de Extremadura, E-06006 Badajoz, Spain}

\begin{abstract}
The Enskog kinetic theory for moderately dense granular suspensions is considered as a model to determine the Navier-Stokes transport coefficients. The influence of the interstitial gas on solid particles is modeled by a viscous drag force term plus a stochastic Langevin-like term. The suspension model is solved by means of the Chapman--Enskog method conveniently adapted to dissipative dynamics. The momentum and heat fluxes as well as the cooling rate are obtained to first order in the deviations of the hydrodynamic field gradients from their values in the homogeneous steady state. Since the cooling terms (arising from collisional dissipation and viscous friction) cannot be compensated for by the energy gained by grains due to collisions with the interstitial gas, the reference distribution (zeroth-order approximation of the Chapman--Enskog solution) depends on time through its dependence on temperature. On the other hand, to simplify the analysis and given that we are interested in computing transport properties in the first order of deviations from the reference state, the steady-state conditions are considered. This simplification allows us to get explicit expressions for the Navier--Stokes transport coefficients. \vicente{The
present work extends previous results [Garz\'o \emph{et al.} 2013, Phys Rev. E \textbf{87}, 032201] since it incorporates two extra ingredients (an additional density dependence of the zeroth-order solution and the density dependence of the reduced friction coefficient) not accounted for by the previous theoretical attempt. While these two new ingredients do not affect the shear viscosity coefficient, the transport coefficients associated with the heat flux as well as the first-order contribution to the cooling rate are different from those obtained in the previous study.} In addition, as expected, the results show that the dependence of the transport coefficients on both inelasticity and density is clearly different from that found in its granular counterpart (no gas phase). Finally, a linear stability analysis of the hydrodynamic equations with respect to the homogeneous steady state is performed. In contrast to the granular case (no gas-phase), no instabilities are found and hence, the homogeneous steady state is (linearly) stable.
\end{abstract}

\draft
\date{\today}
\maketitle

\section{Introduction}
\label{sec1}

Although in nature granular matter is surrounded by an interstitial fluid (like the air, for instance), most of theoretical and computational studies have neglected the impact of the gas phase on the dynamics of solid particles. On the other hand, it is known that in many practical applications (like for instance species segregation in granular mixtures \cite{MLNJ01,NSK03,SSK04,WXZS08,CPSK10,PGM14}) the effect of the surrounding fluid on grains cannot be ignored. Needless to say, at a kinetic theory level, the description of granular suspensions (\vicente{namely, a suspension of solid particles in a viscous gas}) is a quite complex problem since a complete microscopic description of the \vicente{gas-solid} system involves the solution of a set of two coupled kinetic equations for each one of the velocity distribution functions of the different phases. Thus, due to the mathematical difficulties embodied in this approach and in order to gain some insight into this problem, an usual model for describing gas-solid flows \cite{KH01} is to consider a kinetic equation for the solid particles where the influence of the surrounding fluid on them is modeled by means of an effective external force. As usual \cite{GTSH12,HTG17}, the external force modeling the effect of the gas phase is constituted by two terms: (i) a viscous drag force (via a term involving a drift or friction coefficient $\gamma$) accounting for the friction of grains on the interstitial fluid and (ii) a stochastic Langevin-like term (via a term involving the background or bath temperature $T_\text{ex}$) accounting for the energy gained by the grains due to their collisions with particles of the background fluid.

Recently the above suspension model has been employed to study the so-called discontinuous shear thickening in non-Newtonian gas-solid suspensions \cite{HTG17,GG19}. The results show the transition from the discontinuous shear thickening (observed for very dilute gases) to the continuous shear thickening as the density of the system increases. These analytical results (approximately obtained by means of Grad's moment method \cite{HTG17} and from an exact solution of the Boltzmann equation for inelastic Maxwell models \cite{GG19}) compare quite well with molecular dynamics simulations \cite{HTG17} for conditions of practical interest. This good agreement highlights again the good performance of kinetic theory tools in reproducing the transport properties of gas-solid flows.

On the other hand, to the best of our knowledge, most of the efforts in kinetic theory of granular suspensions has been mainly focused on non-Newtonian transport properties (which are directly related with the pressure tensor). In particular, much less is known about the energy transport in gas-solid flows. The knowledge of the transport coefficients associated with the heat flux is interesting by itself and also for possible practical applications in suspensions where temperature and density gradients are present in the system. In this context, it would be desirable to provide simulators with the appropriate expressions of the \vicente{Navier--Stokes} transport coefficients to work when studying gas-solid flows where collisions among particles are inelastic.

The aim of this paper is to determine the Navier--Stokes transport coefficients of granular suspensions in the framework of the Enskog kinetic equation. Since this equation applies for moderate densities \vicente{(let's say for instance, solid volume fraction $\phi \lesssim 0.25$ for hard spheres)}, the comparison between kinetic theory and molecular dynamics simulations becomes practical. Attempts on the evaluation of the Navier--Stokes transport coefficients for granular suspensions \vicente{modeled by the Enskog equation} have been previously published. Thus, in Ref.\ \cite{GTSH12} the authors determined the transport coefficients of gas-solid flows \vicente{starting from the suspension model constituted by the viscous drag force plus the stochastic Langevin term}. Their results show that the effect of the gas phase on both the shear viscosity and the diffusive heat conductivity coefficients is non-negligible for industrially relevant portions of the parameter space. However, for the sake simplicity, the temperature dependence of the scaled friction coefficient $\gamma^*=\gamma/\nu(T)$ (where $\nu\propto T^{1/2}$ is an effective collision frequency for hard spheres and $T$ is the granular temperature) was implicitly neglected in the above calculations \cite{GTSH12} to get analytic (explicit) expressions for the transport coefficients. The above temperature dependence of $\gamma^*$ was accounted for in a subsequent paper \cite{GFHY16} \emph{but} for a simplified model where only the drag force term was considered in the Enskog equation.

\vicente{A more careful study was carried out later in Ref.\ \cite{GChV13} where the transport coefficients were explicitly computed by considering both the temperature dependence of the reduced friction coefficient as well as the complete form of the suspension model. On the other hand, although computer simulations \cite{KS99} have clearly shown that the friction coefficient depends on the volume fraction, the calculations performed in Ref.\ \cite{GChV13} were carried out by assuming that the driven parameters of the model are constant. Needless to say, the impact of the density dependence of $\gamma$ on transport properties is expected to be more relevant as the gas phase becomes denser. Apart from this simplification, although not explicitly stated, another limitation of the above theory \cite{GChV13} is that it was obtained by neglecting contributions to the transport coefficients coming from an additional density dependence of the zeroth-order distribution $f^{(0)}$ (in fact, although this simplification was noted in a subsequent erratum \cite{GChV13bis}, it has not been implemented so far in the calculations). This extra density dependence of $f^{(0)}$ is expected to be involved in the evaluation of the heat flux transport coefficients.}

\vicente{The question arises then as to whether, and if so to what extent, the conclusions drawn from Ref.\ \cite{GChV13} may be altered when the above two new ingredients (density dependence of both the distribution $f^{(0)}$ and the friction coefficient $\gamma$) are accounted for in the theory. In this paper we address this question by extending the results derived in Ref.\ \cite{GChV13} to situations not covered by previous studies on granular suspensions. The present theory subsumes all previous analyses \cite{GTSH12,GFHY16,GChV13}, which are recovered in the appropriate limits. In particular, a comparison between the results obtained here for the transport coefficients with those derived in \cite{GChV13} shows that while the expression of the shear viscosity coefficient is formally equivalent to the one obtained before, the heat flux transport coefficients and the first-order contribution to the cooling rate differ from those reported in Ref.\ \cite{GChV13}.}

As in previous works \cite{GD99a,L05,GTSH12}, the transport coefficients are obtained by solving the Enskog equation by means of the application of the Chapman--Enskog method \cite{CC70}. Since a reference equilibrium state is missing in granular gases, an important point in the Chapman--Enskog expansion is the choice of the zeroth-order solution $f^{(0)}$ (reference base state of the perturbation scheme). While in the dry granular case (no gas phase) the distribution  $f^{(0)}$ is chosen to be the local version of the homogeneous cooling state, there is more flexibility in the choice of $f^{(0)}$ in driven granular gases (\vicente{or, equivalently in gas-solid flows}). In the case of gas-solid flows \cite{GChV13}, for simplicity one possibility is to take a steady distribution $f^{(0)}$ at any point of the system \cite{GM02,G11a}. However, the presence of \vicente{the interstitial fluid} introduces the possibility of a local energy unbalance and hence, the zeroth-order distribution is not in general a stationary distribution. This fact introduces new contributions to the transport coefficients, which were not considered when a local steady state was assumed at zeroth-order \cite{GM02,G11a}. Thus, for \emph{general} small deviations from the homogeneous steady state the energy gained by grains due to collisions with the background fluid cannot be compensated locally with the cooling terms (viscous friction plus inelastic collisions). Thus, although we are interested in determining the transport coefficients under steady state conditions, we have to start from an \emph{unsteady} zeroth-order solution in order to achieve the integral equation verifying the first-order solution $f^{(1)}$. The solution to this equation under steady state conditions provides the explicit forms of the transport coefficients.

The plan of the paper is as follows. In section \ref{sec2}, the Enskog kinetic equation for granular suspensions is introduced and the corresponding balance equations for the densities of mass, momentum, and energy are derived. Then, section \ref{sec3} studies the homogeneous steady state where some theoretical predictions are compared against available computer simulation results. The comparison shows an excellent agreement for conditions of practical interest. Section \ref{sec4} addresses the Chapman--Enskog expansion around the unsteady reference distribution $f^{(0)}(\mathbf{r}, \mathbf{v}, t)$ up to first-order in spatial gradients.  The explicit expressions of the Navier--Stokes transport coefficients and the cooling rate are displayed in section \ref{sec5} for steady state conditions. In dimensionless form, these coefficients are given in terms of the coefficient of restitution $\al$, the volume fraction $\phi$, and the (reduced) background temperature $T_\text{ex}^*$. The dependence of the transport coefficients and the cooling rate on the parameter space is illustrated for several systems showing that the influence of the gas phase on them is in general quite significant. As an application of the results found here, a linear stability analysis of the Navier--Stokes hydrodynamic equations around the homogeneous steady state is carried out in section \ref{sec6}; the analysis shows that the homogeneous state is linearly stable. \vicente{This finding agrees with the previous stability analysis performed in Ref.\ \cite{GChV13}}. We close the paper in section \ref{sec7} with a brief discussion of the results reported here.

\section{Enskog kinetic equation for granular suspensions}
\label{sec2}

We consider a set of solid particles of diameter $\sigma$ and mass $m$ immersed in a viscous gas. Collisions between grains are inelastic and are characterized by a (positive) constant coefficient of normal restitution $\al \leq 1$, where $\al=1$ corresponds to elastic collisions (ordinary gases). At moderate densities, the one-particle velocity distribution function of solid particles  $f(\mathbf{r}, \mathbf{v}; t)$ obeys the Enskog kinetic equation
\beq
\label{2.1}
\frac{\partial f}{\partial t}+\mathbf{v}\cdot \nabla f+\mathcal{F} f=J_\text{E}[\mathbf{r},\mathbf{v}|f,f],
\eeq
where
\beq
\label{2.2}
J_{\text{E}}\left[{\bf r}, {\bf v}_{1}|f,f\right]=\sigma^{d-1}\int d{\bf v}_{2}\int d\widehat{\boldsymbol{\sigma}}\,\Theta (\widehat{{\boldsymbol {\sigma}}}\cdot {\bf g}_{12})(\widehat{\boldsymbol {\sigma }}\cdot {\bf g}_{12})\left[\alpha^{-2}f_2(\mathbf{r},\mathbf{r}-\boldsymbol {\sigma }, \mathbf{v}_1'', \mathbf{v}_2'',t)-f_2(\mathbf{r},\mathbf{r}+\boldsymbol {\sigma}, \mathbf{v}_1, \mathbf{v}_2,t)\right]
\eeq
is the Enskog collision operator. Here,
\beq
\label{2.3}
f_2(\mathbf{r}_1,\mathbf{r}_2, \mathbf{v}_1, \mathbf{v}_2,t)=\chi({\bf r}_1,{\bf r}_2)
f({\bf r}_1, {\bf v}_1, t)f({\bf r}_2, {\bf v}_2, t),
\eeq
$d$ is the dimensionality of
the system ($d=2$ for disks and $d=3$ for spheres), $\boldsymbol
{\sigma}=\sigma \widehat{\boldsymbol {\sigma}}$, $\widehat{\boldsymbol
{\sigma}}$ being a unit vector, $\Theta $ is the Heaviside step function, and ${\bf g}_{12}={\bf v}_{1}-{\bf v}_{2}$. The double primes on the velocities in Eq.\ \eqref{2.2} denote the initial values $\{\mathbf{v}_1'', \mathbf{v}_2''\}$ that lead to $\{\mathbf{v}_1, \mathbf{v}_2\}$ following a binary collision:
\beq
\label{2.4}
{\bf v}_{1}''={\bf v}_{1}-\frac{1}{2}\left( 1+\alpha^{-1}\right)(\widehat{{\boldsymbol {\sigma }}}\cdot {\bf g}_{12})\widehat{\boldsymbol {\sigma}}, \quad
{\bf v}_{2}''={\bf v}_{2}+\frac{1}{2}\left( 1+\alpha^{-1}\right)
(\widehat{{\boldsymbol {\sigma }}}\cdot {\bf g}_{12})\widehat{
\boldsymbol {\sigma}}.
\eeq
In addition, $\chi[{\bf r},{\bf r}\pm\boldsymbol{\sigma}|\{n(t)] $ is the equilibrium pair correlation function at contact as a functional of the nonequilibrium density field $n({\bf r}, t)$ defined by
\begin{equation}
\label{2.5}
n({\bf r}, t)=\int\; d{\bf v} f({\bf r},{\bf v},t).
\end{equation}

In Eq.\ \eqref{2.1}, the operator $\mathcal{F}$ represents the fluid-solid interaction force that models the effect of the viscous gas on solid particles. In order to fully account for the influence of the interstitial molecular fluid on the dynamics of grains, a instantaneous fluid force model is employed \cite{GTSH12,GFHY16,HTG17}. For low Reynolds numbers, it is assumed that the external force $\mathbf{F}$ acting on solid particles is composed by two independent terms. One term corresponds to a viscous drag force $\mathbf{F}^{\text{drag}}$ proportional to the (instantaneous) velocity of particle $\mathbf{v}$. This term takes into account the friction of grains on the viscous gas. Since the model attempts to mimic gas-solid flows, the drag force is defined in terms of the relative velocity $\mathbf{v}-\mathbf{U}_g$ where $\mathbf{U}_g$ is the (known) mean flow velocity of the surrounding molecular gas. Thus, the drag force $\mathbf{F}^{\text{drag}}=-m\gamma \left(\mathbf{v}-\mathbf{U}_g\right)$ is represented in the Enskog equation \eqref{2.1} by the term
\beq
\label{2.6}
\mathcal{F}^{\text{drag}}f\to -\gamma \frac{\partial}{\partial \mathbf{v}}\cdot\left(\mathbf{v}-\mathbf{U}_g\right)f,
\eeq
where $\gamma$ is the drag or friction coefficient. The second term in the total force corresponds to a stochastic force that tries to simulate the kinetic energy gain due to eventual collisions with the (more rapid) molecules of the background fluid. It does this by adding a random velocity to each particle between successive collisions \cite{WM96}. This stochastic force $\mathbf{F}^{\text{st}}$ has the form of a Gaussian white noise with the properties \cite{K81}
\beq
\label{2.7}
\langle \mathbf{F}_i^{\text{st}}(t) \rangle =\mathbf{0}, \quad
\langle \mathbf{F}_i^{\text{st}}(t) \mathbf{F}_j^{\text{st}}(t') \rangle = 2 m \gamma T_\text{ex} \mathsf{I} \delta_{ij}\delta(t-t'),
\eeq
where $\mathsf{I}$ is the unit tensor and $i$ and $j$ refer to two different particles. Here, $T_\text{ex}$ can be interpreted as the temperature of the background (or bath) fluid. In the context of the Enskog kinetic equation, the stochastic external force is represented by a Fokker--Planck operator of the form  \cite{K81,NE98}
\beq
\label{2.7.1}
\mathcal{F}^{\text{st}}f\to -\frac{\gamma T_\text{ex}}{m}\frac{\partial^2 f}{\partial v^2}.
\eeq
Note that the strength of correlation in Eq.\ \eqref{2.7.1} has been chosen to be consistent with the fluctuation-dissipation theorem for elastic collisions \cite{K81}.

Although the drift coefficient $\gamma$ is in general a tensor, here for simplicity  we assume that this coefficient is a scalar proportional to the square root of $T_\text{ex}$ because the drag coefficient is proportional to the viscosity of the solvent \cite{KH01}. In addition, as usual in granular suspension models \cite{K90,KS99}, $\gamma$ is a function of the solid volume fraction
\beq
\label{2.8}
\phi=\frac{\pi^{d/2}}{2^{d-1}d\Gamma \left(\frac{d}{2}\right)}n\sigma^d.
\eeq
Thus, the drift coefficient $\gamma$ can be written as
\beq
\label{2.12}
\gamma=\gamma_0 R(\phi),
\eeq
where $\gamma_0 \propto \eta_g \propto \sqrt{T_\text{ex}}$, $\eta_g$ being the viscosity of the solvent or gas phase.   In the case of hard spheres ($d=3$), for Stokes flow we can use the existing analytical closure derived by Koch \cite{K90} for the function $R(\phi)$ in the case of very dilute suspensions ($\phi \leq 0.1$):
\beq
\label{2.13}
R(\phi)=1+3\sqrt{\frac{\phi}{2}}.
\eeq
For $\phi >0.1$, Koch and Sangani \cite{KS99} used simulations based on multipole expansions to propose the $\phi$-dependence of $R$. It is given by
\beq
\label{2.14}
R(\phi)=1+\frac{3}{\sqrt{2}}\phi^{1/2}+\frac{135}{64}\phi \ln\phi+
11.26\phi (1-5.1\phi+16.57\phi^2-21.77\phi^3)-\phi \chi(\phi)\ln \epsilon_m.
\eeq
Here, $\epsilon_m \sigma$ can be regarded as a length scale characterizing the impact of non-continuum effects on the lubrication forces between two smooth particles at contact. Typical values of $\epsilon_m$ are in the range 0.01--0.05. Since this term contributes to $R(\phi)$ through a weak logarithmic factor, the influence of its explicit value is not important in the final results. Here, we take $\epsilon_m=0.01$ as a typical value.

\vicente{The suspension model defined by Eqs.\ \eqref{2.1}, \eqref{2.6}, and \eqref{2.7.1} is a simplified version of the model employed in Ref.\ \cite{GChV13} to get the Navier--Stokes transport coefficients. In this latter model \cite{GTSH12}, the friction coefficient of the drag force ($\gamma_b$ in the notation  of Ref.\ \cite{GChV13}) and the strength of the correlation ($\xi_b^2$ in the notation  of Ref.\ \cite{GChV13}) are considered to be in general different. Here, as mentioned before, both coefficients are related as $\xi_\text{b}^2=2\gamma_\text{b}T_\text{ex}/m^2$ to be consistent with the fluctuation-dissipation theorem. Thus, some of the results derived in this paper (mainly those regarding homogeneous states) can be directly obtained from those reported in Ref.\ \cite{GChV13} by making the changes  $\gamma_b\to m \gamma$ and $\xi_b^2 \to 2 \gamma T_\text{ex}/m$ with $R(\phi)=1$. We have preferred in this paper to adopt the notation introduced in Eqs.\ \eqref{2.6} and \eqref{2.7.1} because this is the notation used in previous studies of sheared granular suspensions \cite{HTG17,GG19}.}

According to Eqs. \eqref{2.6} and \eqref{2.7.1}, the Enskog equation \eqref{2.1} reads
\beq
\label{2.9}
\frac{\partial f}{\partial t}+\mathbf{v}\cdot \nabla f-\gamma\Delta \mathbf{U}\cdot \frac{\partial f}{\partial \mathbf{v}}-\gamma\frac{\partial}{\partial \mathbf{v}}\cdot \mathbf{V} f-\gamma \frac{T_{\text{ex}}}{m}\frac{\partial^2 f}{\partial v^2}=J_\text{E}[\mathbf{r},\mathbf{V}|f,f].
\eeq
Here, $\Delta \mathbf{U}=\mathbf{U}-\mathbf{U}_g$, $\mathbf{V}=\mathbf{v}-\mathbf{U}$ is the peculiar velocity, and
\beq
\label{2.10}
\mathbf{U}(\mathbf{r},t)=\frac{1}{n(\mathbf{r},t)}\int d\mathbf{v}\; \mathbf{v} f(\mathbf{r},\mathbf{v},t)
\eeq
is the mean particle velocity. Another relevant hydrodynamic field is the \emph{granular} temperature $T(\mathbf{r},t)$ defined as
\beq
\label{2.11}
T(\mathbf{r},t)=\frac{m}{d n(\mathbf{r},t)} \int d\mathbf{v}\; V^2 f(\mathbf{r},\mathbf{v},t).
\eeq
\vicente{Note that in the model defined in \cite{GChV13}) the mean flow velocity of the interstitial gas is assumed to be equal to the mean flow velocity of solid particles ($\mathbf{U}_g=\mathbf{U}$) for the sake of simplicity.}

The macroscopic balance equations for the granular suspension are obtained when one
multiplies the Enskog equation \eqref{2.9}  by $\{1, m{\bf v}, m v^2\}$ and integrates over
velocity. After some algebra, one gets the balance equations \cite{GD99a,GTSH12,GChV13}
\begin{equation}
D_{t}n+n\nabla \cdot {\bf U}=0\;, \label{2.15}
\end{equation}
\begin{equation}
D_{t}{\bf U}=-\rho ^{-1}\nabla \cdot \mathsf{P}-\gamma \Delta \mathbf{U}\;,
\label{2.16}
\end{equation}
\begin{equation}
D_{t}T+\frac{2}{dn} \left( \nabla \cdot {\bf q}+\mathsf{P}:\nabla {\bf U}\right) =2\gamma \left(T_\text{ex}-T\right)-\zeta \,T.
\label{2.17}
\end{equation}
In the above equations, $D_{t}=\partial_{t}+{\bf U}\cdot \nabla$ is the material derivative and $\rho=m n$ is the mass density. The
cooling rate $\zeta$ is proportional to $1-\alpha^2$ and is due to dissipative collisions. The pressure tensor ${\sf P}({\bf r},t)$ and
the heat flux ${\bf q}({\bf r},t)$ have both {\em kinetic} and {\em collisional transfer} contributions, i.e., ${\sf P}={\sf P}^\text{k}+{\sf P}^\text{c}$ and ${\bf q}={\bf q}^\text{k}+{\bf q}^\text{c}$. Their kinetic contributions
are defined by
\begin{equation}
\label{2.18}
{\sf P}^\text{k}=\int \; d{\bf v} m{\bf V}{\bf V}f({\bf r},{\bf v},t), \quad
{\bf q}^\text{k}=\int \; d{\bf v} \frac{m}{2}V^2{\bf V}f({\bf r},{\bf v},t),
\end{equation}
and the collisional transfer contributions are \cite{GD99a}
\beq
{\sf P}^{\text{c}}=\frac{1+\alpha}{4}m \sigma^{d}
\int d\mathbf{v}_{1}\int d\mathbf{v}_{2}\int
d\widehat{\boldsymbol {\sigma }}\,\Theta (\widehat{\boldsymbol
{\sigma }}\cdot
\mathbf{g}_{12})(\widehat{\boldsymbol {\sigma }}\cdot \mathbf{g}_{12})^{2}\widehat{\boldsymbol {\sigma }}\widehat{\boldsymbol {\sigma }}  \int_{0}^{1}dx\; f_2\left[\mathbf{r}-x{\boldsymbol
{\sigma }},\mathbf{r}+(1-x) {\boldsymbol {\sigma }},\mathbf{v}_{1},\mathbf{v}_{2}, t\right],
\label{2.19}
\eeq
\beq
\label{2.20}
{\bf q}^{\text{c}}=\frac{1+\alpha}{4}m \sigma^{d}
\int d\mathbf{v}_{1}\int d\mathbf{v}_{2}\int
d\widehat{\boldsymbol {\sigma }}\,\Theta (\widehat{\boldsymbol
{\sigma }}\cdot
\mathbf{g}_{12})(\widehat{\boldsymbol {\sigma }}\cdot \mathbf{g}_{12})^{2}
 ({\bf G}_{12}\cdot\widehat{\boldsymbol {\sigma }})
\widehat{\boldsymbol {\sigma}}\int_{0}^{1}dx\; f_2\left[\mathbf{r}-
x{\boldsymbol{\sigma}},\mathbf{r}+(1-x)
{\boldsymbol {\sigma}},\mathbf{v}_{1},\mathbf{v}_{2}, t\right],\nonumber\\
\eeq
where ${\bf G}_{12}=\frac{1}{2}({\bf V}_1+{\bf V}_2)$ is the velocity of center of mass. Finally, the cooling rate $\zeta$ is given by
\beq
\zeta =\frac{\left(1-\alpha^{2}\right)}{4dnT} m \sigma^{d-1}\int d\mathbf{v}
_{1}\int d\mathbf{v}_{2}\int d\widehat{\boldsymbol {\sigma }}
\Theta (\widehat{\boldsymbol {\sigma }}\cdot
\mathbf{g}_{12})(\widehat{ \boldsymbol {\sigma }}\cdot
\mathbf{g}_{12})^{3}f_2(\mathbf{r}, \mathbf{r}+\boldsymbol {\sigma
},\mathbf{v}_{1},\mathbf{v}_{2}, t). \label{2.21}
\eeq

\vicente{Before closing this section, it is important to recall the range of validity of the suspension model \eqref{2.9}. As already discussed before \cite{GTSH12}, the assumptions made in the model are relevant to the range of dimensionless physical parameters encountered in a circulating fluidized bed (low Reynolds numbers and moderate densities). A crucial aspect of the model is that the form of the Enskog collision operator $J_\text{E}[\mathbf{r}, \mathbf{v}|f,f]$ is assumed to be the same as for a dry granular gas (i.e., when the influence of the interstitial gas is neglected). This means that the collision dynamics does not contain any parameter of the environmental gas. As it has been noted in several papers \cite{K90,TK95,SMTK96,KH01,WKL03}, the above assumption requires that the mean-free time between collisions is assumed to be much less than the time needed by the fluid forces to significantly affect the dynamics of solid particles. Thus, we expect that the suspension model \eqref{2.3} may be reliable in situations where the gas phase has a weak influence on the motion of grains (solid particles immersed in air, for instance). Of course, this assumption fails for instance in the case of liquid flows (high density) where the presence of fluid must be taken into account in the collision process.}

\section{Homogeneous steady state}
\label{sec3}

\vicente{Before computing the transport coefficients, it is instructive to analyze the homogeneous steady state. This state was widely analyzed in Refs.\ \cite{GChV13,ChVG13}. For homogeneous situations}, the density $n$ and the temperature $T$ are spatially uniform, and with an appropriate selection of the frame of reference, the mean flow velocities vanish ($\mathbf{U}=\mathbf{U}_g=\mathbf{0}$). Consequently, Eq.\ \eqref{2.9} becomes
\beq
\label{3.1}
\frac{\partial f}{\partial t}-\gamma\frac{\partial}{\partial \mathbf{v}}\cdot \mathbf{v} f-\gamma \frac{T_{\text{ex}}}{m}\frac{\partial^2 f}{\partial v^2}=J_\text{E}[\mathbf{V}|f,f],
\eeq
where
\beq
\label{3.2}
J_{\text{E}}\left[f, f\right] =\chi \sigma^{d-1}\int
d\mathbf{v}_{2}\int d\widehat{\boldsymbol {\sigma}}\Theta
(\widehat{\boldsymbol {\sigma}}\cdot \mathbf{g}_{12})(\widehat{
\boldsymbol {\sigma }}\cdot \mathbf{g}_{12})\left[ \alpha^{-2}f(v_{1}'')f(v_{2}'')-f(v_{1})f(v_{2})\right].
\eeq
Here, $\chi$ is the pair correlation function evaluated at the (homogeneous) density $n$. The collision operator \eqref{3.2} can be recognized as the Boltzmann operator for inelastic collisions multiplied by the factor $\chi$. For homogeneous states, the only nontrivial balance equation is that of the temperature \eqref{2.17}:
\beq
\label{3.3}
\partial_t T=2\gamma\left(T_{\text{ex}}-T\right)-\zeta T.
\eeq
As usual, for times longer than the mean free time, one expects that the system achieves a hydrodynamic regime where the distribution $f$ qualifies as a \emph{normal} distribution \cite{CC70} in the sense that $f$ depends on time only through its dependence on the temperature $T$. In this regime, $\partial_t f=(\partial_T f)(\partial_t T)$ and Eq.\ \eqref{3.1} reads
\beq
\label{3.4}
\bigg[2\gamma\left(\theta^{-1}-1\right)-\zeta\bigg]T\frac{\partial f}{\partial T}-\gamma\frac{\partial}{\partial\mathbf{v}}\cdot\mathbf{v}f-\frac{\gamma T_{\text{ex}}}{m}\frac{\partial^2 f}{\partial v^2}=J_{\text{E}}[f,f],
\eeq
where $\theta\equiv T/T_\text{ex}$ and use has been made of Eq.\ \eqref{3.3}. In addition, for homogeneous states, Eq.\ \eqref{2.21}  gives the following form for the cooling rate $\zeta$:
\beq
\label{4.7}
\zeta(t)=\frac{\pi^{(d-1)/2}}{4d\Gamma\left(\frac{d+3}{2}\right)}(1-\al^2)\frac{m\sigma^{d-1}}{nT}\chi\int\; d \mathbf{v}_1
\int\; d \mathbf{v}_2\;  g_{12}^3\; f(\mathbf{v}_1,t)\;f(\mathbf{v}_2,t).
\eeq

For elastic collisions ($\al=1$ and so, $\zeta=0$), as expected Eq.\ \eqref{3.4} admits the solution
\beq
\label{3.5}
f_0(v,t)=n \left(\frac{m}{2\pi T(t)}\right)^{d/2} \exp\left(-\frac{m v^2}{2T(t)}\right)
\eeq
where the temperature obeys the time-dependent equation
\beq
\label{3.6}
\partial_t T =2\gamma\left(T_{\text{ex}}-T\right).
\eeq
The system therefore is in a time-dependent ``equilibrium state'' before reaching the asymptotic steady state where $T=T_{\text{ex}}$. For inelastic collisions, $\zeta \neq 0$ and to date the solution to Eq.\ \eqref{3.4} has not been found.

On the other hand, after a transient stage, the system achieves a \emph{steady} state characterized by the steady temperature $T_\text{s}$. According to Eq.\ \eqref{3.6}, $T_\text{s}$ is given by the condition
\beq
\label{3.7}
%\Lambda_\text{s}\equiv
2\gamma \left(\theta_{\text{s}}^{-1}-1\right)-\zeta_{\text{s}}=0,
\eeq
where the subscript $\text{s}$ means that the quantities are evaluated at $T=T_{\text{s}}$. \vicente{At a given value of the environmental temperature $T_\text{ex}$ (which acts as a bath temperature in the sense that it is considered as a thermal energy reservoir), Eq.\ \eqref{3.7} implies that in the steady state the energy gained by grains due to their collisions with the interstitial fluid ($\gamma T_\text{ex}$) is exactly compensated by the cooling terms arising from collisional dissipation ($\zeta T$) and viscous friction ($\gamma T$). Moreover, as usual in the granular literature, the effects of the energy balance on the internal degrees of freedom of grains are not considered in the description.}

\vicente{As shown in previous works \cite{GChV13,ChVG13,GMT12,GMT13}}, dimensionless analysis requires that $f_\text{s}$ has the scaled form
\beq
\label{3.8}
f_{\text{s}}(\mathbf{v},\gamma, T_\text{ex})=n v_0^{-d}\varphi_{\text{s}}(\mathbf{c},\gamma_{\text{s}}^*)\equiv n v_0^{-d}\varphi_{\text{s}}(\mathbf{c},\lambda,\theta_{\text{s}}),
\eeq
where $v_0=\sqrt{2T_\text{s}/m}$ is the thermal speed and the unknown scaled distribution $\varphi_{\text{s}}$ is a function of the dimensionless parameters $\mathbf{c}\equiv \mathbf{v}/v_0$ and $\gamma_{\text{s}}^*$ where
\begin{equation}
\label{3.9}
\gamma_{\text{s}}^*(\lambda,\theta_\text{s})=\lambda\theta_{\text{s}}^{-1/2}, \quad \lambda(\phi)=\frac{\gamma_0 R(\phi)\ell}{\sqrt{2T_{\text{ex}}/m}}=\frac{\sqrt{2}\pi^{d/2}}
{2^d d \Gamma\left(\frac{d}{2}\right)}\frac{R(\phi)}{\phi \sqrt{T_\text{ex}^*}}.
\end{equation}
Here, $T_\text{ex}^*\equiv T_\text{ex}/(m \sigma^2 \gamma_0^2)$ is the (reduced) background gas temperature. In the second relation of Eq.\ \eqref{3.9}, $\ell=1/(n \sigma^{d-1})$ is proportional to the mean free path of hard spheres. \vicente{The scaling given by Eq.\ \eqref{3.8} is equivalent to the one proposed in Refs.\ \cite{GChV13,ChVG13} when one makes the mapping $\xi_\text{s}^*\to 2\lambda \theta_\text{s}^{-3/2}$ with $R(\phi)=1$. Here, $\xi_\text{s}^*$ is defined by Eq.\ (24) of \cite{GChV13}. This means that the results for homogeneous states can be directly obtained from those derived in Refs.\ \cite{GChV13,ChVG13} by making the above change. On the other hand, we have preferred here to revisit the homogeneous state in order to check the previous results.}

In terms of $\varphi_{\text{s}}$, in the steady state, Eq.\ \eqref{3.1} for $f_\text{s}$ can be rewritten as
\begin{equation}
\label{3.10}
-\gamma^*_{\text{s}}\frac{\partial}{\partial\mathbf{c}}\cdot\mathbf{c}\varphi_{\text{s}}-\frac{\gamma^*_{\text{s}}}{2\theta_{\text{s}}}
\frac{\partial^2 \varphi_{\text{s}}}{\partial c^2}=J_{\text{E}}^*[\varphi_{\text{s}},\varphi_{\text{s}}],
\end{equation}
where we have introduced the dimensionless collision operator $J_{\text{E}}^*=\ell v_0^{d-1}J_{\text{E}}/n$. Although the exact form of $\varphi_{\text{s}}$ is not known, an indirect information on it can be obtained from the kurtosis or fourth cumulant
\begin{equation}
\label{3.11}
a_{2,\text{s}}=\frac{4}{d(d+2)}\int\; d{\bf c}\; c^4 \varphi_s(c)-1.
\end{equation}
The cumulant $a_{2,\text{s}}$ measures the deviation of $\varphi_s$ from its Maxwellian form $\pi^{-d/2}e^{-c^2}$. This coefficient can be obtained by multiplying Eq.\ \eqref{3.10} by $c^4$ and integrating over velocity. The result is
\beq
\label{3.11.1}
d(d+2)\left(\gamma_\text{s}^* a_{2,\text{s}}-\frac{1}{2}\zeta_\text{s}^*\right)=\beta_4,
\eeq
where $\zeta_\text{s}^*\equiv \ell \zeta_\text{s}/v_0$ and
\beq
\label{4.11}
\beta_4=\int d\mathbf{c}\; c^4\; J_{\text{E}}^*[\varphi_\text{s},\varphi_\text{s}].
\eeq
Upon deriving Eq.\ \eqref{3.11.1} use has been made again of the steady state condition \eqref{3.7}.

As expected, Eq.\ \eqref{3.11.1} cannot be solved unless one knows the collisional moments $\zeta_\text{s}^*$ and $\beta_4$. As in previous works \cite{NE98,GChV13,ChVG13}, a good estimate of $\zeta_\text{s}^*$ and $\beta_4$ can be obtained by replacing $\varphi_\text{s}$ by its leading Sonine approximation \cite{NE98}:
\beq
\label{3.12}
\varphi_\text{s}\simeq  \frac{e^{-c^2}}{\pi^{d/2}}\left\{1 +a_{2,\text{s}}\left[ \frac{c^4}{2}-\frac{(d+2)c^2}{2}+\frac{d(d+2)}{8}\right]\right\}.
\eeq
In this case, retaining only linear terms in $a_{2,\text{s}}$, one has
\beq
\label{3.12.1}
\zeta_\text{s}^*\to \zeta_0^{(0)}+\zeta_0^{(1)}a_{2,\text{s}}, \quad \beta_4\to \beta_4^{(0)}+\beta_4^{(1)}a_{2,\text{s}},
\eeq
where \cite{NE98}
\beq
\label{4.12}
\zeta_0^{(0)}=\frac{2K}{d}\chi (1-\al^2), \quad \zeta_0^{(1)}=\frac{3}{16}\zeta_0^{(0)},
\eeq
\beq
\label{4.13}
\beta_4^{(0)}=-K \chi \left(1-\alpha^2\right)\left(d+\frac{3}{2}+\alpha^2\right), \quad
\beta_4^{(1)}=-K\chi\left(1-\alpha^2\right)\left[\frac{3}{32}\left(10d+39+10\alpha^2\right)+\frac{d-1}{1-\alpha}
\right],
\eeq
and
\beq
\label{3.13.1}
K=\frac{\pi^{(d-1)/2}}{\sqrt{2}\Gamma(d/2)}.
\eeq
With these results, Eq.\ \eqref{3.11.1} can be easily solved with the result
\beq
\label{3.13}
a_{2,\text{s}}=\frac{16(1-\alpha)(1-2\alpha^2)}{73+56d-3\alpha(35+8d)+30(1-\alpha)\alpha^2
+32d(d+2)\gamma_\text{s}^*/K\chi(1+\alpha)}.
\end{equation}
Notice that in Eq.\ \eqref{3.13}, $\gamma_\text{s}^*$ is consistently obtained from the steady state condition \eqref{3.7} by replacing $\zeta_\text{s}^*\to \zeta_0^{(0)}$. \vicente{The expression \eqref{3.13} agrees with the one obtained in Ref.\ \cite{GChV13} when one takes the steady state condition $\xi_\text{s}^*=2\gamma_\text{s}^*+\zeta_0^{(0)}$  in Eq.\ (31) of \cite{GChV13}.}

\begin{figure}
{\includegraphics[width=0.4 \columnwidth,angle=0]{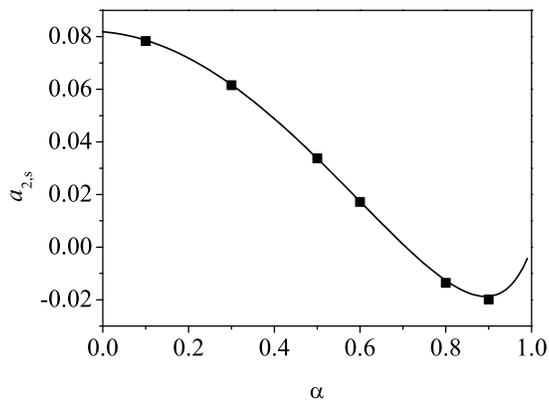}}
\caption{Plot of the fourth cumulant $a_{2,\text{s}}$ as a function of the coefficient of restitution $\al$ for a two-dimensional ($d=2$) granular suspension with $\phi=0.25$. The line is the theoretical result given by Eq.\ \eqref{3.13} (with $R(\phi)=1$) and the symbols are the Monte Carlo simulation results obtained in Ref.\ \cite{GChV13}. The parameters of the simulation are $m=1$, $\sigma=0.01$, $\gamma_0=1$, and $T_\text{ex}=1$.
\label{fig1}}
\end{figure}
\begin{figure}
{\includegraphics[width=0.4 \columnwidth,angle=0]{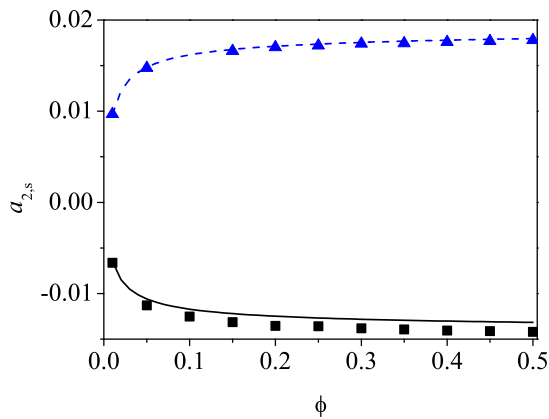}}
\caption{Plot of the fourth cumulant $a_{2,\text{s}}$ as a function of the volume fraction $\phi$ for a two-dimensional ($d=2$) granular suspension. Two different values of the coefficient of restitution are considered: $\al=0.8$ (solid line and squares) and $\al=0.6$ (dashed line and triangles). The lines are the theoretical results given by Eq.\ \eqref{3.13} (with $R(\phi)=1$) and the symbols are the Monte Carlo simulation results. The parameters of the simulation are $m=1$, $\sigma=0.01$, $\gamma_0=1$, and $T_\text{ex}=1$.
\label{fig2}}
\end{figure}
\begin{figure}
{\includegraphics[width=0.4 \columnwidth,angle=0]{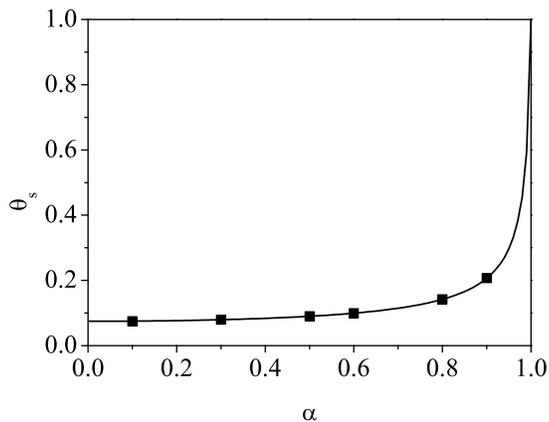}}
\caption{Plot of the (reduced) temperature $\theta_\text{s}\equiv T_\text{s}/T_\text{ex}$  as a function of the coefficient of restitution $\al$ for a two-dimensional ($d=2$) granular suspension with $\phi=0.25$. The line is the theoretical result given by Eq.\ \eqref{3.16} (with $R(\phi)=1$) and the symbols are the Monte Carlo simulation results obtained in Ref.\ \cite{GChV13}.  The parameters of the simulation are $m=1$, $\sigma=0.01$, $\gamma_0=1$, and $T_\text{ex}=1$.
\label{fig3}}
\end{figure}
\begin{figure}
{\includegraphics[width=0.4 \columnwidth,angle=0]{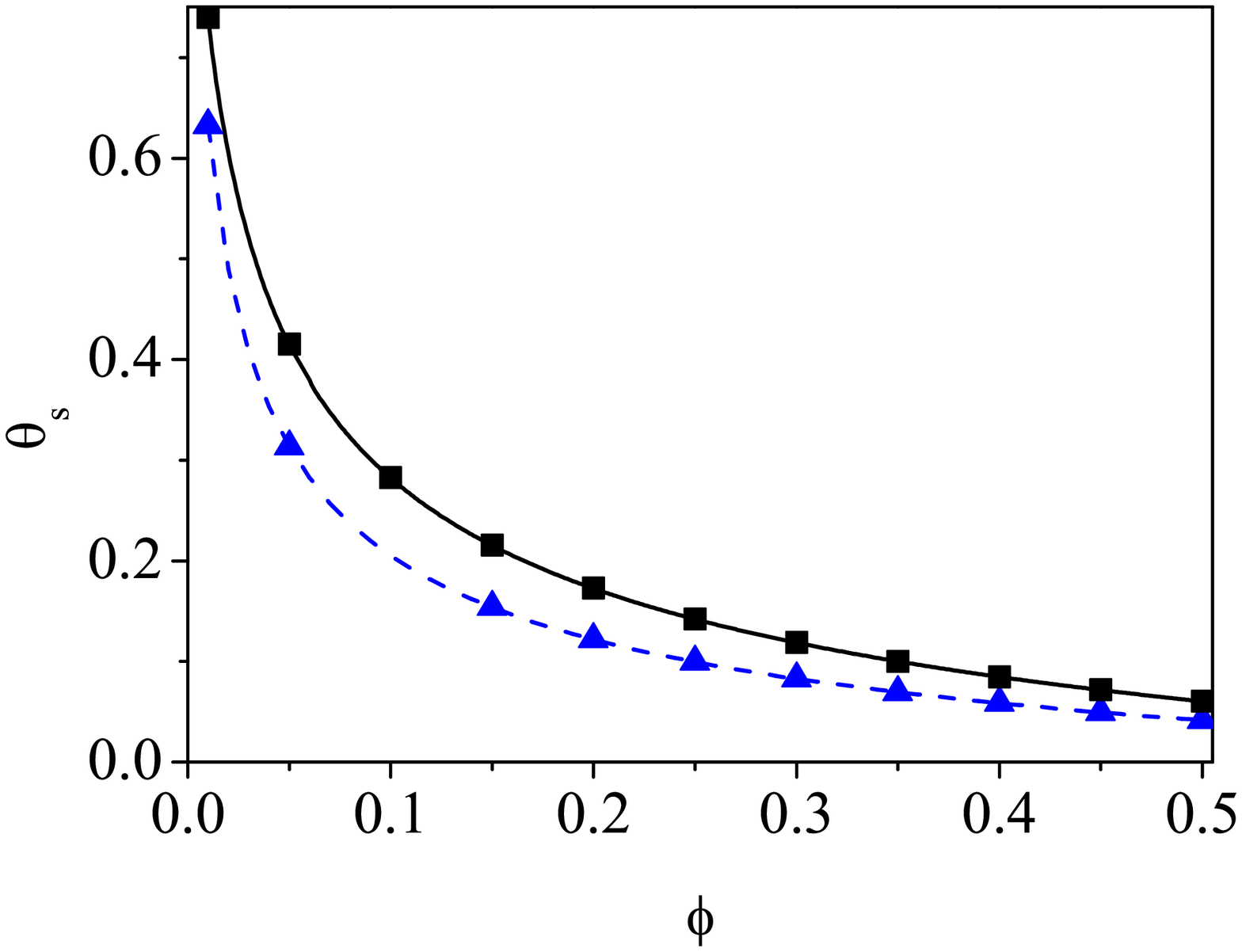}}
\caption{Plot of the (reduced) temperature $\theta_\text{s}\equiv T_\text{s}/T_\text{ex}$ as a function of the volume fraction $\phi$ for a two-dimensional ($d=2$) granular suspension. Two different values of the coefficient of restitution are considered: $\al=0.8$ (solid line and squares) and $\al=0.6$ (dashed line and triangles). The lines are the theoretical results given by Eq.\ \eqref{3.16} (with $R(\phi)=1$) and the symbols are the Monte Carlo simulation results.  The parameters of the simulation are $m=1$, $\sigma=0.01$, $\gamma_0=1$, and $T_\text{ex}=1$.
\label{fig4}}
\end{figure}

Once $a_{2,\text{s}}$ is known, the dependence of the cooling rate on both the coefficient of restitution $\al$ and the (reduced) external temperature $T_\text{ex}^*$ can be obtained from the first relation of Eq.\ \eqref{3.12.1}. Finally, the (reduced) steady temperature $\theta_s$ is determined by solving the cubic equation
\beq
\label{3.16}
2\lambda \left(\theta_\text{s}^{-1}-1\right)=\frac{\sqrt{2}}{d}\frac{\pi^{\left(d-1\right)/2}}
{\Gamma \left( \frac{d}{2}\right)}(1-\alpha^2)\chi(\phi) \left(1+\frac{3}{16}a_{2,\text{s}}\right)\sqrt{\theta_\text{s}}.
\eeq
\vicente{As expected, Eq.\ \eqref{3.16} is consistent with Eq.\ (33) of Ref.\ \cite{GChV13} for the steady temperature when one takes $R(\phi)=1$ and makes the replacement $\xi_\text{s}^*\to 2\lambda \theta_\text{s}^{-3/2}$.}

Figure \ref{fig1} shows the $\al$-dependence of the fourth cumulant $a_{2,\text{s}}$ for hard disks ($d=2$) with the solid volume fraction $\phi=0.25$. In the case of hard disks, we have chosen the following form for $\chi(\phi)$ \cite{T95}:
\beq
\label{3.17}
\chi(\phi)=\frac{1-\frac{7}{16}\phi}{(1-\phi)^2}.
\eeq
The theoretical results given by Eq.\ \eqref{3.13} are compared against the results obtained in Ref.\ \cite{GChV13} by numerically solving the Enskog equation from the direct simulation Monte Carlo (DSMC) method \cite{B94}. The parameters of the simulation are $m=1$, $\sigma=0.01$, $\gamma_0=1$, and $T_\text{ex}=1$. In addition, the function $R(\phi)=1$ in the simulations. \vicente{Although this figure was already presented in Ref.\ \cite{GChV13}, we plot it again here to remark the} excellent agreement between theory and simulations observed in the complete range of values of $\al$. Since the values of $a_{2,\text{s}}$ are very small (in fact their magnitude is smaller than the one found in the dry granular case \cite{NE98, MS00}) then, the Sonine approximation \eqref{3.12} can be considered as a good representation of the scaled distribution $\varphi_s(c)$. As a complement of Fig.\ \ref{fig1}, Fig.\ \ref{fig2} shows $a_{2,\text{s}}$ versus $\phi$ for two values of $\al$. It is quite apparent that the qualitative dependence of the fourth cumulant on the density depends strongly on the inelasticity since while $a_{2,\text{s}}$ decreases monotonically with $\phi$ at $\al=0.8$, the opposite happens at $\al=0.6$. \vicente{We do not actually have an intuitive explanation for the change of behaviour of $a_{2,\text{s}}$ when the coefficient of restitution varies from 0.8 to 0.6.} Next, the (reduced) temperature $\theta_s$ is considered. Figure \ref{fig3} shows $\theta_\text{s}$ versus $\al$ for $d=2$, $\phi=0.25$, and the same parameters as the one considered in Figs.\ \ref{fig1} and \ref{fig2}. First, as expected, $\theta_\text{s}=1$ for elastic collisions. Moreover, the steady granular temperature decreases with inelasticity. It is illustrated in Fig.\ \ref{fig4} (\vicente{which was also plotted in Ref.\ \cite{GChV13}}) where $\theta_\text{s}$ is plotted against the density $\phi$ for two different values of $\al$. Figures \ref{fig3} and \ref{fig4} highlight again the excellent agreement between theory and simulations, even for extreme values of both inelasticity and/or density.

\section{Transport around the homogeneous steady state. Chapman--Enskog expansion}
\label{sec4}

As in previous studies \cite{GD99a,GChV13,KG13}, we assume that we perturb the homogeneous steady state by small spatial gradients. These perturbations give rise to nonzero contributions to the pressure tensor and the heat flux, which are characterized by transport coefficients. The evaluation of the transport coefficients is the main objective of the present contribution. In order to get them, we will solve the Enskog equation \eqref{2.9} by means of the Chapman--Enskog method \cite{CC70} conveniently adapted to granular fluids. As usual, the Chapman--Enskog method assumes the existence of a normal solution such that all space and time dependence of the velocity distribution function occurs through the hydrodynamic fields, namely,
\begin{equation}
f({\bf r},{\bf v},t)=f\left[{\bf v}|n (t), T(t), {\bf U}(t) \right].
\label{4.1}
\end{equation}
The notation on the right hand side indicates a functional dependence on the density, temperature and flow velocity. For small spatial variations (i.e., low Knudsen numbers), this functional dependence can be made local in space through an expansion in the gradients of the hydrodynamic fields. To generate it, $f$ is written as a series expansion in a formal parameter $\epsilon$ measuring the non-uniformity of the system,
\begin{equation}
f=f^{(0)}+\epsilon \,f^{(1)}+\epsilon^2\,f^{(2)}+\cdots, \label{4.2}
\end{equation}
where each factor of $\epsilon$ means an implicit gradient of a hydrodynamic field. In contrast to the case of dry granular gases \cite{GD99a}, in ordering the different level of approximations in the kinetic equation, one has to characterize the magnitude of the drift term $\gamma$ relative to the gradients as well as the term $\Delta \mathbf{U}$. With respect to the first term, since $\gamma$ does not induce any flux in the system, it is considered to be of zeroth-order in gradients. Regarding the term $\Delta \mathbf{U}$, since in the absence of gradients $\mathbf{U}$ tends to $\mathbf{U}_g$ after a transient period, then $\Delta \mathbf{U}$ is expected to be at least to first order in the spatial gradients.

According to the expansion \eqref{4.2}, the Enskog operator $J_\text{E}$ and the time derivative $\partial_t$ are also given in the representations
\beq
\label{4.3}
J_{\text{E}}=J_{\text{E}}^{(0)}+\epsilon J_{\text{E}}^{(1)}+\cdots, \quad \partial_t=\partial_t^{(0)}+\epsilon
\partial_t^{(1)}+\cdots .
\eeq
The coefficients in the time derivative expansion are identified by a representation of the fluxes and the cooling rate in the macroscopic balance equations as a similar series through their definitions as functionals of $f$. This is the usual Chapman--Enskog method \cite{CC70,GS03} for solving kinetic equations. The expansions \eqref{4.3} yield similar expansions for the heat and momentum fluxes and the cooling rate when substituted into Eqs.\ \eqref{2.18}--\eqref{2.21}:
\begin{equation}
\label{4.4}
P_{ij}=P_{ij}^{(0)}+\epsilon P_{ij}^{(1)}+\cdots, \quad  {\bf q}={\bf q}^{(0)}+\epsilon {\bf q}^{(1)}+\cdots, \quad
\zeta=\zeta^{(0)}+\epsilon \zeta^{(1)}+\cdots.
\end{equation}
Here, we shall restrict our calculations to the first order in the uniformity parameter $\epsilon$.

\subsection{Zeroth-order approximation}

To zeroth order in the expansion, the distribution $f^{(0)}$ obeys the kinetic equation
\beq
\label{4.5}
\partial_t^{(0)}f^{(0)}-\gamma\frac{\partial}{\partial \mathbf{v}}\cdot \mathbf{V} f^{(0)}-\gamma \frac{T_{\text{ex}}}{m}\frac{\partial^2 f^{(0)}}{\partial v^2}=J_\text{E}^{(0)}[f^{(0)},f^{(0)}],
\eeq
where $J_{\text{E}}^{(0)}[f^{(0)},f^{(0)}]$ is given by Eq.\ \eqref{3.2} with the replacement $f_\text{s}\to f^{(0)}({\bf r}, {\bf v},t)$. The conservation laws at this order are given by $\partial_t^{(0)}n=0$, $\partial_t^{(0)}\mathbf{U}=\mathbf{0}$, and
\beq
\label{4.6}
\partial_t^{(0)}T=2\gamma\left(T_{\text{ex}}-T\right)-\zeta^{(0)} T,
\eeq
where $\zeta^{(0)}$ is determined from Eq.\ \eqref{2.21} to zeroth order. In particular, as said in section \ref{sec3}, a good approximation to
$\zeta^{(0)}$ is given by the first relation of Eq.\ \eqref{3.12.1}, namely,
\beq
\label{4.6.1}
\zeta^{(0)}=\frac{2}{d}\frac{\pi^{\left( d-1\right) /2}}
{\Gamma \left( \frac{d}{2}\right)}(1-\alpha^2)\chi \left(1+\frac{3}{16}a_{2}\right)
n \sigma^{d-1}\sqrt{\frac{T}{m}}.
\eeq

The kinetic equation \eqref{4.5} can be rewritten in terms of the derivative $\partial_T f^{(0)}$ when one takes into account the zeroth-order balance equations:
\beq
\label{4.7.1}
\left[2\gamma\left(\theta^{-1}-1\right)-\zeta^{(0)}\right]T\frac{\partial f^{(0)}}{\partial T}-\gamma\frac{\partial}{\partial\mathbf{v}}\cdot\mathbf{V}f^{(0)}-\frac{\gamma T_{\text{ex}}}{m}\frac{\partial^2 f^{(0)}}{\partial v^2}=J_{\text{E}}^{(0)}[f^{(0)},f^{(0)}].
\eeq
Equation \eqref{4.7.1} has the same form as the corresponding Enskog equation \eqref{3.4} for a strictly homogeneous state. However, in Eq.\ \eqref{4.7.1} $f^{(0)}(\mathbf{r}, \mathbf{v}, t)$ is a \emph{local} distribution. Therefore, as in the homogeneous state, the solution to Eq.\ \eqref{4.7.1} can be written in the form \eqref{3.8} (with the replacement $T_\text{s}\to T$) where the scaled distribution $\varphi(\mathbf{c},\lambda,\theta)$ obeys the \emph{unsteady} equation
\beq
\label{4.8}
\left[2\gamma^*\left(\theta^{-1}-1\right)-\zeta_0^*\right]\theta\frac{\partial\varphi}{\partial\theta}
+
\left(\frac{\zeta_0^*}{2}-\gamma^*\theta^{-1}\right)\frac{\partial}{\partial \mathbf{c}}\cdot \mathbf{c}\varphi-\frac{\gamma^*}{2\theta}\frac{\partial^2 \varphi}{\partial c^2}=J_{\text{E}}^*[\varphi,\varphi],
\eeq
where $\zeta_0^*\equiv \ell \zeta^{(0)}/v_0(T)$ and $\gamma^*=\lambda \theta^{-1/2}$. Upon deriving Eq.\ \eqref{4.8} use has been made of the property
\beq
\label{4.9}
T\frac{\partial f^{(0)}}{\partial T}=-\frac{1}{2}\frac{\partial}{\partial \mathbf{V}}\cdot\mathbf{V}f^{(0)}+nv_0^{-d}\theta\frac{\partial \varphi}{\partial \theta},
\eeq
where the derivative $\partial \varphi/\partial \theta$ is taken at constant $\mathbf{c}$.

The velocity distribution function $f^{(0)}$ is isotropic in $\mathbf{V}$ so that, according to Eqs.\ \eqref{2.18}--\eqref{2.20}, the heat flux to zeroth-order vanishes as expected ($\mathbf{q}^{(0)}=\mathbf{0}$) and the pressure tensor $P_{ij}^{(0)}=p\delta_{ij}$, where the hydrostatic pressure is
\beq
\label{4.9.1}
p=n T \left[1+2^{d-2}(1+\al)\phi \chi \right].
\eeq

\begin{figure}
{\includegraphics[width=0.4 \columnwidth,angle=0]{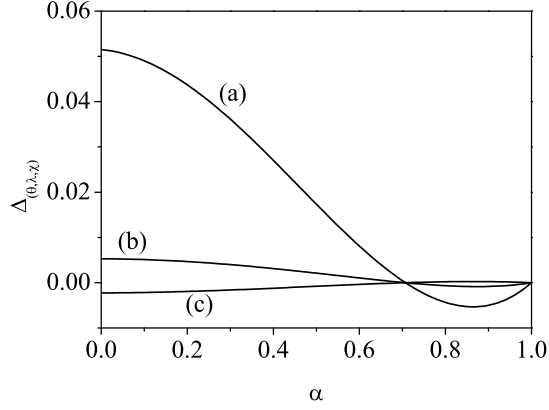}}
\caption{Plot of the derivatives $\Delta_\theta$ (a), $\Delta_\lambda$ (b), and $\Delta_\chi$ (c) for $d=3$, $\phi=0.25$, and $T_\text{ex}^*=0.9$.
\label{fig5}}
\end{figure}

As discussed in section \ref{sec3}, although the explicit form of $\varphi$ is not known, a good approximation is given by the Sonine approximation \eqref{3.12}. In particular, the equation for the \emph{unsteady} fourth cumulant $a_2$ can be easily obtained from Eq.\ \eqref{4.8} as
\beq
\label{4.10}
\frac{d(d+2)}{4}\Lambda^{(0)}\theta\frac{\partial a_2}{\partial\theta}
+d\left(d+2\right)\left(\gamma^*\theta^{-1}-\frac{\zeta_0^*}{2}\right)\left(1+a_2\right)-d\left(d+2\right)\gamma^*\theta^{-1}=\beta_4,
\eeq
where $\Lambda^{(0)}\equiv 2\gamma^*\left(\theta^{-1}-1\right)-\zeta_0^*$ and $\beta_4$ is defined in Eq.\ \eqref{4.11}. In the steady state, $\Lambda^{(0)}=0$ and the solution to Eq.\ \eqref{4.10} is given by Eq.\ \eqref{3.13} once one expands $\zeta_0^*$ and $\beta_4$ in powers of $a_2$. Beyond the steady state, Eq.\ \eqref{4.10} must be numerically solved to get the dependence of $a_2$ on the (reduced) temperature. On the other hand, as we will show in section \ref{sec5}, in order to get the transport coefficients in the steady state we need to know the derivatives $\Delta_\theta\equiv (\partial a_2/\partial \theta)_\text{s}$, $\Delta_\lambda\equiv (\partial a_2/\partial \lambda)_\text{s}$, and $\Delta_\chi\equiv (\partial a_2/\partial \chi)_\text{s}$. These derivatives provide an indirect information (through the fourth cumulant) on the departure of the time-dependent solution $f^{(0)}$ from its stationary form $f_\text{s}$. According to Eq.\ \eqref{4.10}, the former derivative is given by
\beq
\label{4.13.1}
\frac{\partial a_2}{\partial \theta}=\frac{\frac{4}{d(d+2)}\beta_4^{(0)}+2\zeta_0^{(0)}+2\left(\frac{2}{d(d+2)}\beta_4^{(1)}-2\gamma^* \theta^{-1}+\frac{19}{16}\zeta_0^{(0)}\right)a_2}{\theta\left[2\gamma^*(\theta^{-1}-1)-
\left(\zeta_0^{(0)}+\zeta_0^{(1)}a_2\right)\right]},
\eeq
where here the expansions \eqref{3.12.1} have been considered and as usual nonlinear terms in $a_2$ have been neglected. In the steady state, the numerator and denominator of Eq.\ \eqref{4.13.1} vanish, hence
the quantity $\Delta_\theta$ becomes indeterminate. As in Ref.\ \cite{GChV13}, this problem can be solved by applying
l'H\^opital's rule. The final result yields a quadratic equation for $\Delta_\theta$. However, given that the magnitude of $\Delta_\theta$ is quite small, one can neglect the term proportional to $\Delta_\theta^2$ in the above quadratic equation and obtain the simple expression
\beq
\label{4.14}
\Delta_\theta=\frac{6\gamma_\text{s}^*\theta_{\text{s}}^{-2}a_{2,\text{s}}}
{2\gamma_\text{s}^*-\frac{15}{8}\zeta_0^{(0)}-\frac{4}{d(d+2)}\beta_4^{(1)}}.
\eeq
\vicente{Equation \eqref{4.14} is consistent with Eq.\ (A6) of Ref.\ \cite{GChV13} when one neglects nonlinear terms in $(\partial a_2/\partial \xi^*)_\text{s}$ and takes $\beta=\frac{1}{2}$.} The derivatives $\Delta_\lambda$ and $\Delta_\chi$ can be easily derived from Eq.\ \eqref{4.10} with the results
\beq
\label{4.15}
\Delta_\lambda=\frac{4\theta_{\text{s}}^{-3/2} a_{2,\text{s}}+2\theta_{\text{s}}^{1/2}\left(\theta_{\text{s}}^{-1}-1\right)}
{\frac{4}{d(d+2)}\beta_4^{(1)}-4\gamma_\text{s}^*+\frac{3}{8}\zeta_0^{(0)}},
\eeq
\beq
\label{4.16}
\Delta_\chi=\frac{\frac{4}{d(d+2)}\beta_4^{(0)}+2\zeta_0^{(0)}+\frac{4}{d(d+2)}\beta_4^{(1)}+\frac{19}{8}\zeta_0^{(0)}+\zeta_0^{(0)}\theta_\text{s}
\Delta_\theta}{2\chi\left(2\gamma_\text{s}^*-\frac{2}{d(d+2)}\beta_4^{(1)}-\frac{3}{16}\zeta_0^{(0)}\right)}.
\eeq
Note that in Eqs.\ \eqref{4.14}--\eqref{4.16}, $\theta_{\text{s}}$ is obtained from Eq.\ \eqref{3.16} by neglecting $a_{2,\text{s}}$. The dependence of the derivatives $\Delta_\theta$, $\Delta_\lambda$, and $\Delta_\chi$ on the coefficient of restitution $\al$ is plotted in Fig.\    \ref{fig5} for $d=3$ and $\phi=0.25$. Here, $T_\text{ex}^*=0.9$; this is a typical value for the (reduced) background temperature used in previous simulations \cite{HTG17}. It is seen that while the magnitude of $\Delta_\lambda$, and $\Delta_\chi$ is much smaller than that of the kurtosis $a_{2,\text{s}}$, $\Delta_\theta$ is of the same order of magnitude as $a_{2,\text{s}}$.

\subsection{First-order approximation}

\vicente{The mathematical steps involved in the derivation of the first-order distribution function $f^{(1)}$ are quite similar to those carried out in Ref.\ \cite{GChV13}. On the other hand, given that the calculations performed in this paper take into account some additional density dependencies not accounted for in the previous derivation \cite{GChV13}, we have preferred here to perform an independent calculation where most of the technical details are provided in the Appendix \ref{appA} for the sake of completeness}. To first-order in spatial gradients, $f^{(1)}$ is given by
\beq
\label{4.16.1}
f^{(1)}(\mathbf{V})=\boldsymbol{\mathcal{A}}(\mathbf{V})\cdot\nabla\ln T+\boldsymbol{\mathcal{B}}(\mathbf{V})\cdot\nabla\ln n + \mathcal{C}_{ij}(\mathbf{V})\frac{1}{2}\left(\frac{\partial U_i}{\partial r_j}+\frac{\partial U_j}{\partial r_i}-\frac{2}{d}\delta_{ij}\nabla\cdot\mathbf{U}\right)+\mathcal{D}(\mathbf{V})\nabla\cdot\mathbf{U},
\eeq
where, in the steady state ($\Lambda^{0)}=0$), the quantities $\boldsymbol{\mathcal{A}}$, $\boldsymbol{\mathcal{B}}$, $\mathcal{C}_{ij}$, and $\mathcal{D}$ verify the following set of coupled linear integral equations:
\beq
\label{4.17}
-\left(2\gamma\theta^{-1}
+\frac{1}{2}\zeta^{(0)}+\zeta^{(0)}\theta \frac{\partial\ln \zeta^*_0}{\partial\theta}\right)
\boldsymbol{\mathcal{A}}-\gamma\frac{\partial}{\partial\mathbf{v}}\cdot\mathbf{V}\boldsymbol{\mathcal{A}}-\frac{\gamma T_{\text{ex}}}{m}\frac{\partial^2}{\partial v^2}\boldsymbol{\mathcal{A}}+\mathcal{L}\boldsymbol{\mathcal{A}}=\mathbf{A},
\eeq
\beqa
\label{4.18}
-\gamma\frac{\partial}{\partial\mathbf{v}}\cdot\mathbf{V}\boldsymbol{\mathcal{B}}-\frac{\gamma T_{\text{ex}}}{m}\frac{\partial^2}{\partial v^2}\boldsymbol{\mathcal{B}}+\mathcal{L}\boldsymbol{\mathcal{B}}&=&\mathbf{B}+\left[\zeta^{(0)}\left(1+\phi
\frac{\partial\ln\chi}{\partial\phi}\right)
+\chi\phi\frac{\partial\chi}{\partial\phi}\frac{\partial}{\partial\chi}\left(\frac{\zeta^{(0)}}{\chi}\right)\right.\nonumber\\
& &\left.-\lambda
\left(1-\phi\frac{\partial\ln R}{\partial \phi}\right)\frac{\partial \zeta^{(0)}}{\partial \lambda}
-2\gamma\left(\theta^{-1}-1\right)\phi \frac{\partial\ln R}{\partial \phi}\right]\boldsymbol{\mathcal{A}},
\eeqa
\beq
\label{4.19}
-\gamma\frac{\partial}{\partial\mathbf{v}}\cdot\mathbf{V}\mathcal{C}_{ij}-\frac{\gamma T_{\text{ex}}}{m}\frac{\partial^2}{\partial v^2}\mathcal{C}_{ij}+\mathcal{L}\mathcal{C}_{ij}=C_{ij},
\eeq
\beq
\label{4.20}
-\gamma\frac{\partial}{\partial\mathbf{v}}\cdot\mathbf{V}\mathcal{D}-\frac{\gamma T_{\text{ex}}}{m}\frac{\partial^2}{\partial v^2}\mathcal{D}-\zeta_1^{(1)}T\frac{\partial f^{(0)}}{\partial T} +\mathcal{L}\mathcal{D}=D.
\eeq
\vicente{In Eq.\ \eqref{4.20}, $\zeta_1^{(1)}$ is a functional of $\mathcal{D}$ defined by Eq.\ \eqref{b16}}. Moreover,
in Eqs.\ \eqref{4.17}--\eqref{4.20}, $\mathcal{L}$ is the linearized collision operator
\beq
\label{4.21}
\mathcal{L}f^{(1)}=-\left(J_{\text{E}}^{(0)}[f^{(0)},f^{(1)}]+J_{\text{E}}^{(0)}[f^{(1)},f^{(0)}]\right),
\eeq
$R$ is defined by Eqs.\ \eqref{2.12}--\eqref{2.14} and the coefficients $\mathbf{A}$, $\mathbf{B}$, $C_{ij}$, and $D$ are functions of the peculiar velocity $\mathbf{V}$ and the hydrodynamic gradients. They are defined by Eqs.\ \eqref{a8}--\eqref{a11}. Note that all the quantities appearing in Eqs.\ \eqref{4.17}--\eqref{4.20} are evaluated in the steady state (the subscript $\text{s}$ has been omitted here for the sake of simplicity). Thus, the transport coefficients obtained by solving Eqs.\ \eqref{2.12}--\eqref{2.14} will be provided in terms of the steady temperature $T_\text{s}$. It is worthwhile to remark that since we are here interested in obtaining the momentum and heat fluxes in the first order of the deviations from the steady state, we only need to know the transport coefficients to zeroth order in the deviations. This means that the solution to the integral equations \eqref{4.17}--\eqref{4.20} will provide us the forms of the transport coefficients and the cooling rate in steady state conditions.

\vicente{According to the Chapman--Enskog scheme \cite{CC70}, acceptable solutions to Eqs.\ \eqref{4.17}--\eqref{4.20} must obey
\beq
\label{4.22}
\int d\mathbf{v} \left(1, \mathbf{V}, V^2\right)f^{(1)}=\left(0,\mathbf{0}, 0\right).
\eeq
These are necessary conditions for the solution to the integral equations to exist (the so-called Fredholm alternative \cite{MM56}). Since $\boldsymbol{\mathcal{A}}(\mathbf{V})\propto \mathbf{A}(\mathbf{V})$, $\boldsymbol{\mathcal {B}}(\mathbf{V})\propto \mathbf{B}(\mathbf{V})$, $\mathcal{C}_{ij}(\mathbf{V})\propto C_{ij}(\mathbf{V})$, and $\mathcal{D}(\mathbf{V})\propto D(\mathbf{V})$, then the solubility conditions \eqref{4.22} can be proved when one takes into account the explicit forms of $\mathbf{A}$, $\mathbf{B}$, $C_{ij}$, and $D$.}

\section{Navier--Stokes transport coefficients}
\label{sec5}

To first order in the spatial gradients, the constitutive equations for the pressure tensor $P_{ij}^{(1)}$ and the heat flux $\mathbf{q}^{(1)}$ are
\beq
\label{5.1}
P_{ij}^{(1)}=-\eta \left(\frac{\partial U_i}{\partial r_j}+\frac{\partial U_j}{\partial r_i}-\frac{2}{d}\delta_{ij}\nabla\cdot\mathbf{U}\right)
-\eta_\text{b}\delta_{ij}\nabla \cdot \mathbf{U},
\eeq
\beq
\label{5.2}
\mathbf{q}^{(1)}=-\kappa \nabla T-\mu \nabla n.
\eeq
Here, $\eta$ is the shear viscosity, $\eta_\text{b}$ is the bulk viscosity, $\kappa$ is the thermal conductivity, and $\mu$ is the diffusive heat conductivity. This latter coefficient vanishes for ordinary gases ($\al=1$). While the coefficients $\eta$, $\kappa$, and $\mu$ have kinetic and collisional contributions, the bulk viscosity $\eta_\text{b}$ has only collisional contributions and hence, it vanishes for dilute gases. In addition, as already mentioned in Ref. \cite{GTSH12}, the forms of the collisional contributions to the transport coefficients are exactly the same as those obtained in the dry granular case (namely, in the absence of the gas phase) \cite{GD99a,L05}, except that $a_{2,\text{s}}$ depends on $\gamma^*$. Thus, we will focus here our attention on the kinetic contributions to the transport coefficients and the cooling rate. Some technical details on this calculation are provided in the Appendix \ref{appB}.

\subsection{Shear and bulk viscosities}

The bulk viscosity $\eta_\text{b}$ is given by
\beq
\label{5.3}
\eta_\text{b}=\frac{2^{2d+1}}{\pi(d+2)}\phi^2 \chi (1+\alpha)\left(1-\frac{a_{2,\text{s}}}{16} \right)\eta_0,
\eeq
where
\begin{equation}
\label{5.4}
\eta_0=\frac{d+2}{8}\frac{\Gamma \left( \frac{d}{2}\right)}{\pi ^{\left( d-1\right) /2}}\sigma^{1-d}\sqrt{mT_\text{s}}
\end{equation}
is the low density value of the shear viscosity for an ordinary gas of hard spheres ($\al=1$). The shear viscosity $\eta$ can be written as
\beq
\label{5.5}
\eta=\frac{\eta_0}{\nu_{\eta}^*+2K'\gamma_\text{s}^*}\left[1-\frac{2^{d-2}}{d+2}\chi\phi(1+\alpha)(1-3\alpha)\right]
\left[1+\frac{2^{d-1}}{d+2}(1+\alpha)\phi\chi\right]+\frac{d}{d+2}\eta_\text{b},
\eeq
where $K'=(d+2)/8K$, $K$ is defined by Eq.\ \eqref{3.13.1} and the (reduced) collision frequency $\nu_\eta^*$ is \cite{GSM07}
\begin{equation}
\label{5.5.1}
\nu_{\eta}^*=\frac{3}{4d}\chi\left(1-\alpha+\frac{2}{3}d\right)(1+\alpha)\left(1+\frac{7}{16}a_{2,\text{s}}\right),
\end{equation}
where $a_{2,\text{s}}$ is defined by Eq.\ \eqref{3.13}. \vicente{The expression \eqref{5.5} for the shear viscosity agrees with the one obtained in Ref.\ \cite{GChV13} when $R(\phi)=1$. This is because the new contributions to the fluxes coming from the extra density dependencies not accounted for in \cite{GChV13} do not affect the form of the pressure tensor.}

\subsection{Thermal conductivity and diffusive heat conductivity}

The thermal conductivity is given by
\beq
\label{5.6}
\kappa=\kappa_{\text{k}}\left[1+3\frac{2^{2-d}}{d+2}\phi\chi\left(1+\alpha\right)\right]+\frac{2^{2d+1}\left(d-1\right)}
{\left(d+2\right)^2\pi}\phi^2\chi\left(1+\alpha\right)\left(1+\frac{7}{16}a_{2,\text{s}}\right)\kappa_0,
\eeq
where
\beq
\label{5.7}
\kappa_0=\frac{d(d+2)}{2(d-1)}\frac{\eta_0}{m}
\eeq
is the low density value of the thermal conductivity for an ordinary gas of hard spheres ($\al=1$) and $\kappa_{\text{k}}$ denotes the kinetic contribution to the thermal conductivity. Its explicit expression is
\beq
\label{5.8}
\kappa_{\text{k}}=\kappa_0\frac{d-1}{d}\frac{1+2a_{2,\text{s}}+\theta_\text{s}\Delta_\theta+3\frac{2^{d-3}}{d+2}\chi\phi\left(1+\alpha\right)^2
\left[2\alpha-1+a_{2,\text{s}}\left(1+\alpha\right)+\frac{1}{2}\left(1+\alpha\right)\theta_\text{s}\Delta_\theta\right]}
{\nu_{\kappa}^*+K'\left(\gamma_\text{s}^*-\frac{3}{2}\zeta_0^*-\theta_\text{s}\zeta_0^{(1)}\Delta_\theta\right)},
\eeq
where $\zeta_0^{(1)}$ is defined by Eq.\ \eqref{4.12} and the derivative $\Delta_\theta$ is given by Eq.\ \eqref{4.14}. In addition, the (reduced) collision frequency $\nu_\kappa^*$ is \cite{GSM07}
\beq
\label{5.9}
\nu_{\kappa}^*=\frac{1+\alpha}{d}\chi\left[\frac{d-1}{2}+\frac{3}{16}\left(d+8\right)\left(1-\alpha\right)
+\frac{296+217d-3\left(160+11d\right)\alpha}{256}a_{2,\text{s}}\right].
\eeq
\vicente{To compare the expression \eqref{5.8} with the one derived in Ref.\ \cite{GChV13} (see Eq.\ (65) of this reference), one has to make the mapping $\xi_\text{s}^*(\partial a_2/\partial \xi^*)_\text{s} \to -(2/3)\theta_\text{s} \Delta_\theta$ and takes $R(\phi)=1$. In this case, we find that the form \eqref{5.8} of the thermal conductivity coefficient is consistent with the one obtained in \cite{GChV13} except for the last term of the numerator (i.e., the term proportional to $\frac{1}{2}\left(1+\alpha\right)\theta_\text{s}\Delta_\theta$). This term comes from the collision integral \eqref{b13}. We have checked that Eq.\ \eqref{5.8} gives the correct result and hence it fixes the slight mistake of Eq.\ (65) of Ref.\ \cite{GChV13}.}

The diffusive heat conductivity $\mu$ is
\beq
\label{5.10}
\mu=\mu_\text{k}\left[1+3\frac{2^{d-2}}{d+2}\phi\chi\left(1+\alpha\right)\right],
\eeq
where the kinetic contribution $\mu_\text{k}$ is given by
\beqa
\label{5.11}
\mu_\text{k}&=&\frac{\kappa_0T_\text{s}}{n}\left(\nu_\kappa^*-3K'\gamma_\text{s}^*\right)^{-1}\Bigg\{\frac{\kappa_{\text{k}}}{\kappa_0}\Bigg[
K' \zeta_0^*\left(1+\phi\frac{\partial\ln\chi}{\partial\phi}\right)+K' \zeta_0^{(1)}\Bigg(\phi\frac{\partial\chi}{\partial\phi} \Delta_\chi-
\lambda\left(1-\phi \frac{\partial \ln R}{\partial \phi}\right)\Delta_\lambda\Bigg)\nonumber\\
& & -2\left(\theta_\text{s}^{-1}-1\right)\gamma_\text{s}^* \phi \frac{\partial \ln R}{\partial \phi}\Bigg]+
\frac{d-1}{d}\Bigg[a_{2,\text{s}}-\lambda\left(1-\phi \frac{\partial \ln R}{\partial \phi}\right)\Delta_\lambda+
\phi\frac{\partial\chi}{\partial\phi}\Delta_\chi\Bigg] \nonumber\\
& &
+3\frac{2^{d-4}\left(d-1\right)}{d\left(d+2\right)}\chi\phi\left(1+\alpha\right)^3
\Bigg[\phi\frac{\partial\chi}{\partial\phi}\Delta_\chi-\lambda\left(1-\phi \frac{\partial \ln R}{\partial \phi}\right)\Delta_\lambda\Bigg]
\nonumber\\
& & +3\frac{2^{d-2}\left(d-1\right)}{d\left(d+2\right)}\chi\phi\left(1+\alpha\right)\left(1+\frac{1}{2}\phi\frac{\partial\ln\chi}{\partial\phi}\right)
\Big[\alpha\left(\alpha-1\right)+\frac{a_{2,\text{s}}}{6}\left(10+2d-3\alpha+3\alpha^2\right)\Big]\Bigg\}.
\eeqa
Here, the derivatives $\Delta_\lambda$ and $\Delta_\chi$ are defined by Eqs.\ \eqref{4.15} and \eqref{4.16}, respectively. \vicente{The expression \eqref{5.11} agrees with Eq.\ (69) of Ref.\ \cite{GChV13}} when one neglects (i) the density dependence of the function $R$ (i.e., $\partial_\phi R=0$) and (ii) all the derivatives of $a_2$ with respect to $\theta$, $\lambda$, and $\chi$ in the steady state (i.e., $\Delta_\theta=\Delta_\lambda=\Delta_\chi=0$). In addition, as in the case of a dry granular gas \cite{BDKS98,GD99a,L05}, the coefficient $\mu$ vanishes for elastic collisions.

\subsection{Cooling rate}

The cooling rate is
\beq
\label{5.12}
\zeta=\zeta_\text{s}^{(0)}+\zeta_U \nabla \cdot \mathbf{U},
\eeq
where $\zeta_\text{s}^{(0)}$ is given by Eq.\ \eqref{4.6.1} with the replacement $T\to T_\text{s}$. The coefficient $\zeta_U$ can be written as
\beq
\label{5.13}
\zeta_U=\zeta_1^{(0)}+\zeta_1^{(1)},
\eeq
where
\beq
\label{5.14}
\zeta_1^{(0)}=-3\frac{2^{d-2}}{d}\chi \phi (1-\al^2),
\eeq
\beqa
\label{5.15}
\zeta_1^{(1)}&=&\frac{9(d+2)2^{d-8}}{d^2}\chi\left(1-\alpha^2\right)\Big(\nu_{\gamma}^*+4K'\gamma_\text{s}^*\Big)^{-1}
\Bigg\{\frac{\omega^*\phi\chi}{2(d+2)}-2^{2-d}\frac{d}{3}\Big[\lambda\left(1-\phi \frac{\partial \ln R}{\partial \phi}\right)\Delta_\lambda
\nonumber\\
& & -\phi\frac{\partial\chi}{\partial\phi}\Delta_\chi-\frac{2}{d}\theta_\text{s}\Delta_\theta \Big]
-\left(1+\alpha\right)\left(\frac{1}{3}-\alpha\right)\left(2a_{2,\text{s}}
+\theta_\text{s}\Delta_\theta\right)\phi\chi\Bigg\}.
\eeqa
Here, we have introduced the quantities
\beq
\label{5.16}
\omega^*=\left(1+\alpha\right)\left\{\left(1-\alpha^2\right)\left(5\alpha-1\right)-\frac{a_{2,\text{s}}}{6}
\left[15\alpha^3-3\alpha^2+3\left(4d+15\right)\alpha-\left(20d+1\right)\right]\right\},
\eeq
\beq
\label{5.17}
\nu_{\gamma}^*=-\frac{1+\alpha}{192}\chi \left[30\alpha^3-30\alpha^2+\left(105+24d\right)\alpha-56d-73\right].
\eeq
It is quite apparent that $\zeta_U=0$ for elastic collisions ($\al=1$). \vicente{As in the case of the diffusive heat conductivity, to compare Eq.\ \eqref{5.15} with the expression (73) for $\zeta_1^{(1)}$ obtained in Ref.\ \cite{GChV13} one has to make the replacement $\theta \Delta_\theta \to -(3/2)\xi_\text{s}^*(\partial a_2/\partial \xi^*)_\text{s}$, take $R(\phi)=1$, and neglect the derivatives of $a_2$ with respect to $\lambda$ and $\chi$ ($\Delta_\lambda=\Delta_\chi=0$). After these changes, we see that both results agree except for a misprint we have found in Eq.\ (73) of Ref.\ \cite{GChV13}.} Note also that $\zeta_U \neq 0$ for dilute granular suspensions \cite{GMT13}.

\subsection{Some illustrative examples}

In summary, the Navier--Stokes transport coefficients $\eta_\text{b}$, $\eta$, $\kappa$, and $\mu$ are given by Eqs.\ \eqref{5.3}, \eqref{5.5}, \eqref{5.6}, and \eqref{5.10}, respectively, while the first-order contribution $\zeta_U$ to the cooling rate is given by Eqs.\ \eqref{5.13}--\eqref{5.15}. As expected, all these coefficients present an intricate dependence on the coefficient of restitution $\al$, the density $\phi$, and the (reduced) background temperature $T_\text{ex}^*$. In addition, their dimensionless forms are defined in terms of the steady temperature $\theta_\text{s}$ and the derivatives $\Delta_\theta$, $\Delta_\lambda$, and $\Delta_\chi$. While these derivatives are explicitly given by Eqs.\ \eqref{4.14}--\eqref{4.16}, the granular temperature is given in terms of the physical solution of the cubic equation \eqref{3.16}.

\begin{figure}
{\includegraphics[width=0.4 \columnwidth,angle=0]{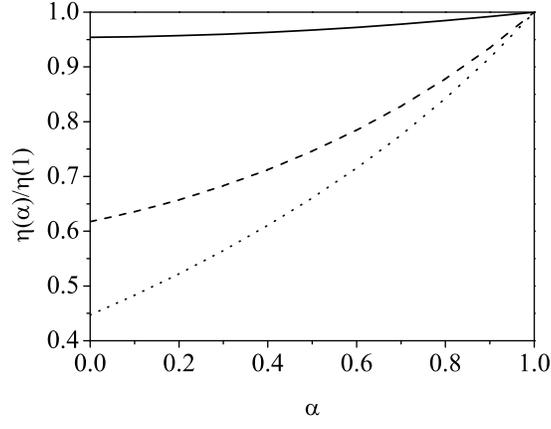}}
\caption{Dependence of the (scaled) shear viscosity $\eta(\al)/\eta(1)$ on the coefficient of restitution $\al$ for $d=3$, $T_\text{ex}^*=0.9$, and three different values of the solid volume fraction: $\phi=0.01$ (a), $\phi=0.1$ (b), and $\phi=0.2$ (c). Here, $\eta(1)$ refers to the shear viscosity coefficient of a suspension with elastic collisions.
\label{fig6}}
\end{figure}
\begin{figure}
{\includegraphics[width=0.4 \columnwidth,angle=0]{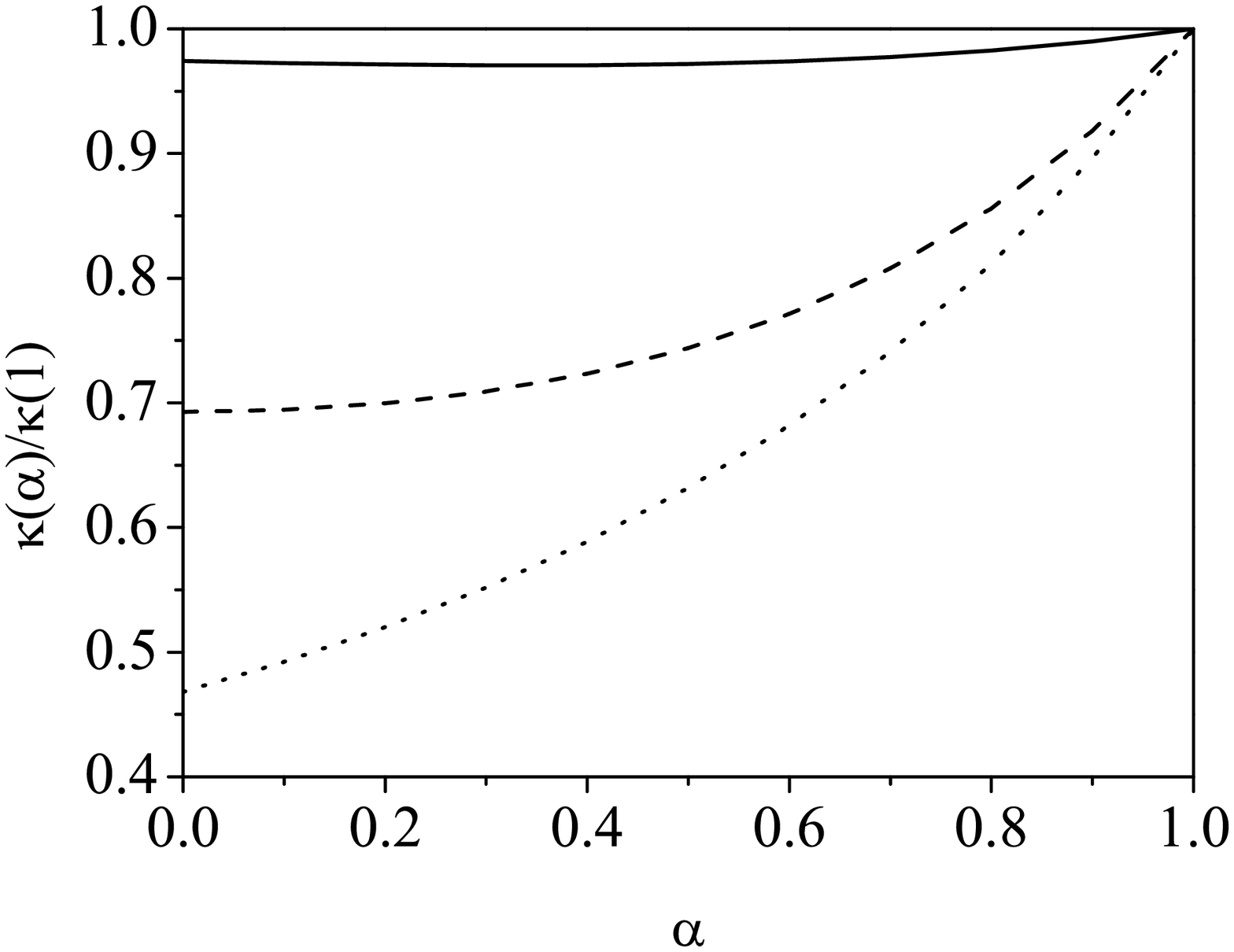}}
\caption{Dependence of the (scaled) thermal conductivity $\kappa(\al)/\kappa(1)$ on the coefficient of restitution $\al$ for $d=3$, $T_\text{ex}^*=0.9$, and three different values of the solid volume fraction: $\phi=0.01$ (a), $\phi=0.1$ (b), and $\phi=0.2$ (c). Here, $\kappa(1)$ refers to the thermal conductivity coefficient of a suspension with elastic collisions.
\label{fig7}}
\end{figure}
\begin{figure}
{\includegraphics[width=0.4 \columnwidth,angle=0]{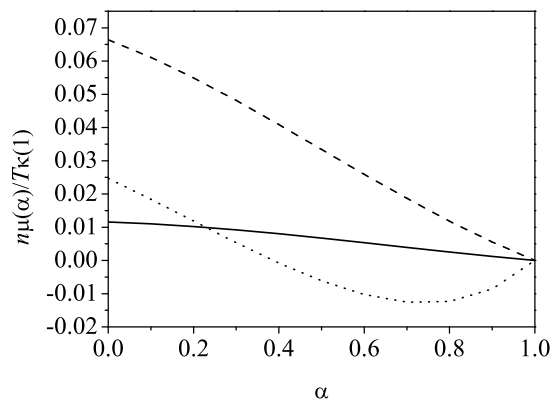}}
\caption{Dependence of the (scaled) diffusive heat conductivity $n\mu(\al)/T\kappa(1)$ on the coefficient of restitution $\al$ for $d=3$, $T_\text{ex}^*=0.9$, and three different values of the solid volume fraction: $\phi=0.01$ (a), $\phi=0.1$ (b), and $\phi=0.2$ (c). Here, $\kappa(1)$ refers to the thermal conductivity coefficient of a suspension with elastic collisions.
\label{fig8}}
\end{figure}

\vicente{As in previous works \cite{GTSH12,GChV13}}, it is quite apparent that one of the principal new features of the present paper lies on the dependence of the Navier--Stokes transport coefficients of granular suspensions on the coefficient of restitution $\al$. Therefore, to illustrate the differences between granular ($\al\neq 1$) and ordinary ($\al=1$) suspensions, the transport coefficients are scaled with respect to their values for elastic collisions. In addition, we consider a three-dimensional system ($d=3$) with $T_\text{ex}^*=0.9$ and three different values of the volume fraction $\phi$: $\phi=0.01$ (very dilute system), $\phi=0.1$, and $\phi=0.2$ (moderately dense system).

In Figs.\ \ref{fig6}--\ref{fig8}, the above Navier--Stokes transport coefficients are plotted as functions of $\al$. While in the case of the shear viscosity and thermal conductivity coefficients we observe that their deviation from their forms for elastic collisions is in general significant, no happens the same in the case of the diffusive heat conductivity since the magnitude of the scaled coefficient $n \mu(\al)/T \kappa(1)$ is much smaller than that of the (scaled) coefficient $\kappa(\al)/\kappa(1)$. Since both $\kappa$ and $\mu$ characterize the heat flux, one could neglect the term proportional to the density gradient in the heat flux. Thus, for practical purposes and analogously to ordinary
(elastic) suspensions, one could assume that the heat flux verifies Fourier's law $\mathbf{q}^{(1)}=-\kappa \nabla T$. With respect to the $\al$-dependence of $\eta$ and $\kappa$, Figures \ref{fig6} and \ref{fig7} highlight that both transport coefficients are \vicente{decreasing} functions of the inelasticity regardless of the density of the system. In addition, the influence of collisional dissipation on momentum and heat transport increases with density, being very tiny in the limit of dilute suspensions. A comparison with the results obtained for dry granular fluids (see for instance, Fig.\ 1 of Ref.\ \cite{G05}) shows significant differences between dry (no gas phase) and granular suspensions. In particular, both theory \cite{GD99a,L05,G13} and simulations \cite{MSG05} show that for relatively dilute dry granular gases ($\phi \lesssim 0.1$) $\eta$ increases with inelasticity, while the opposite occurs for sufficiently dense dry granular fluids ($\phi \gtrsim 0.1$). The same qualitative behavior is observed for the thermal conductivity coefficient \cite{GD99a,L05,G13}. This non-monotonic behavior contrasts with the predictions found here for granular suspensions where $\eta$ and $\kappa$ always decreases with decreasing $\al$. Regarding the coefficient $\mu$, we see that the impact of density on it is significant since while $\mu$ is always positive for dilute suspensions, it can be negative for moderately dense suspensions. It is worthwhile to note that the behavior of the shear viscosity and thermal conductivity on both density and coefficient of restitution found here is qualitatively similar to that of a confined quasi-two-dimensional granular fluid \cite{GBS18}.

\begin{figure}
\includegraphics[width=0.4 \columnwidth,angle=0]{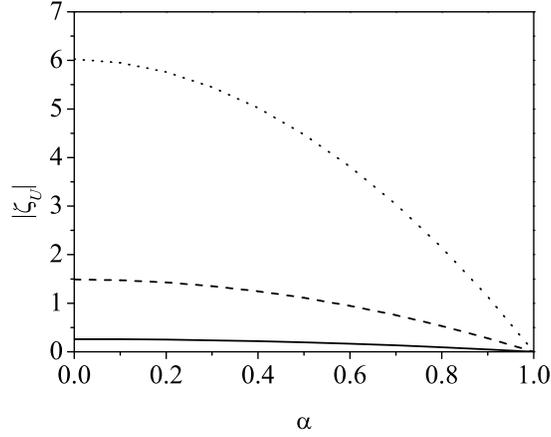}
\caption{Dependence of the magnitude of the first-order contribution $|\zeta_U|$ to the cooling rate on the coefficient of restitution $\al$ for $d=3$, $T_\text{ex}^*=0.9$, and three different values of the solid volume fraction: $\phi=0.1$ (a), $\phi=0.3$ (b), and $\phi=0.5$ (c).
\label{fig9}}
\end{figure}

Finally, the dependence of the magnitude of the first-order contribution $|\zeta_U|$ to the cooling rate is plotted in Fig.\ \ref{fig9} for the same parameters employed in Figs.\ \ref{fig6}--\ref{fig8}. As the coefficient $\mu$, $\zeta_U=0$ for elastic collisions. On the other hand, in contrast to the diffusive heat conductivity, we observe that the influence of inelasticity on $\zeta_U$ is important, specially at large densities. This means that the contribution of  $\zeta_U$ to the cooling rate must be considered as the inelasticity increases.

\section{Stability of the homogeneous steady state}
\label{sec6}

The knowledge of the Navier--Stokes transport coefficients and the cooling rate opens up the possibility of solving the hydrodynamic equations for $n$, $\mathbf{U}$, and $T$ for situations close to the homogeneous steady state. The solution of the linearized hydrodynamic equations allows us to study the stability of the homogeneous steady state. This is likely one of the nicest applications of the Navier--Stokes equations. In order to obtain them, one has to substitute the equation of state \eqref{4.9.1}, the Navier--Stokes constitutive equations \eqref{5.1} and \eqref{5.2} for the pressure tensor and heat flux, respectively, and Eq.\ \eqref{5.12} for the cooling rate into the exact balance equations \eqref{2.15}--\eqref{2.17}. The Navier--Stokes hydrodynamic equations read
\begin{equation}
\label{6.1}
D_t n+\nabla\cdot\mathbf{U}=0,
\end{equation}
\begin{equation}
\label{6.2}
D_t U_i+\rho^{-1}\partial_i p=\rho^{-1}\partial_j\left[\eta\left(\partial_iU_j+\partial_jU_i-\frac{2}{d}\delta_{ij}
\nabla\cdot\mathbf{U}\right)+\eta_\text{b}\delta_{ij}\nabla\cdot\mathbf{U}\right]-\gamma \Delta \mathbf{U},
\end{equation}
\beqa
\label{6.3}
\Big(D_t+2\gamma\left(1-\theta^{-1}\right)+\zeta^{(0)}\Big)T&=&\frac{2}{dn}\nabla\cdot\left(\kappa\nabla T+\mu\nabla n\right)+\frac{2}{dn}\bigg[\eta\left(\partial_iU_j+\partial_jU_i-\frac{2}{d}\delta_{ij}\nabla\cdot\mathbf{U}\right)\nonumber\\
& & +
\eta_\text{b}\delta_{ij}\nabla\cdot\mathbf{U}\bigg]\partial_iU_j-T\zeta_U\nabla\cdot\mathbf{U}-\frac{2}{dn}p\nabla\cdot\mathbf{U}.
\eeqa
As mentioned in several previous papers \cite{G05,GMD06}, the general form of the cooling rate $\zeta$ should include second-order gradient contributions of the form $\zeta_n \nabla^2 n$ and $\zeta_T \nabla^2 T$ in Eq.\ \eqref{6.3}. Nevertheless, as shown for a dilute (dry) granular gas \cite{BDKS98}, given that the ratios $\zeta_n/\mu$ and $\zeta_T/\kappa$ were shown to be very small for not very inelastic particles, the terms $\zeta_n \nabla^2 n$ and $\zeta_T \nabla^2 T$ were neglected in the Navier--Stokes transport equations. We assume that the same happens for dense gases and hence, these second-order contributions can be neglected for practical purposes. Apart from this approximation, Eqs.\ \eqref{6.1}--\eqref{6.3} are exact to second order in the spatial gradients for a granular suspension at moderate densities.

\vicente{The stability analysis of the homogeneous steady state was also carried out in Ref.\ \cite{GChV13}. On the other hand and as mentioned in section \ref{sec1}, the present work generalizes the results derived before \cite{GChV13} since it takes into account both an extra density dependence of the zeroth-order distribution $f^{(0)}$ and the dependence of the friction coefficient $\gamma$ on the volume fraction $\phi$ ($R(\phi)\neq 1$). Thus, it is worth to assess to what extent the previous theoretical results \cite{GChV13} are indicative of what happens in the stability analysis of the homogeneous state when the above density dependencies for the transport coefficients and the cooling rate are considered. This is the main motivation of this Section.}

To analyze the stability of the homogeneous solution, Eqs.\ \eqref{6.1}--\eqref{6.3} must be linearized around the homogeneous steady state. In this state, the hydrodynamic fields take the homogeneous steady values $n\equiv\text{const.}$, $T_\text{s}\equiv\text{const.}$, and $\mathbf{U}_g=\mathbf{U}\equiv \mathbf{0}$. For small spatial gradients, we assume that the deviations $\delta y_{\beta}(\mathbf{r},t)=y_{\beta}(\mathbf{r},t)-y_{\beta,\text{s}}$ are small, where $\delta y_{\beta}(\mathbf{r},t)$ denotes the deviations of the hydrodynamic fields $\left\{y_\beta; \beta=1,\cdots, d+2\right\}=\left\{n, \mathbf{U}, T\right\}$ from their values in the homogeneous \textit{steady} state. Moreover, as usual we also suppose that the interstitial fluid is not perturbed and hence, $\mathbf{U}_g=\mathbf{U}=\mathbf{0}$.

It must be recalled that here, in contrast to the linear stability analysis for dry granular gases \cite{G05,BP04,G19}, the reference state is stationary and so one does not have to eliminate the time dependence of the transport coefficients. On the other hand, in order to compare our results with those obtained for granular fluids \cite{G05}, the following space and time variables are introduced:
\beq
\label{6.4}
\tau=\frac{1}{2}n \sigma^{d-1}\sqrt{\frac{T_\text{s}}{m}}t,\quad \mathbf{r}'=\frac{1}{2}n \sigma^{d-1}\mathbf{r}.
\eeq
The dimensionless time scale $\tau$ measures the average number of collisions per particle in the time interval between 0 and $t$. The unit length $\mathbf{r}'$ is proportional to the mean free path of solid particles. As usual, a set of Fourier transformed dimensionless variables are then introduced by
\begin{equation}
\label{6.5}
\rho_{\mathbf{k}}(\tau)=\frac{\delta n_{\mathbf{k}}(\tau)}{n},\quad\mathbf{w}_{\mathbf{k}}(\tau)=\frac{\delta\mathbf{U}_{\mathbf{k}}(\tau)}{\sqrt{T_\text{s}/m}},
\quad\theta_{\mathbf{k}}(\tau)=\frac{\delta T_{\mathbf{k}}(\tau)}{T_\text{s}},
\end{equation}
where $\delta y_{\mathbf{k}\beta}\equiv\left\{\rho_{\mathbf{k}}(\tau),\mathbf{w}_{\mathbf{k}}(\tau),\theta_{\mathbf{k}}(\tau)\right\}$ is defined as
\beq
\label{6.6}
\delta y_{\mathbf{k}\beta}(\tau)=\int d\mathbf{r}'\text{e}^{-i\mathbf{k}\cdot\mathbf{r}'}\delta y_{\beta}(\mathbf{r}',\tau),
\eeq
where here the wave vector $\mathbf{k}$ is dimensionless.

In terms of the above dimensionless variables, as expected, the $d-1$ transverse velocity components
$\mathbf{w}_{\mathbf{k}\perp}=\mathbf{w}_{\mathbf{k}}-\left(\mathbf{w}_{\mathbf{k}}\cdot\widehat{\mathbf{k}}\right)\widehat{\mathbf{k}}$ (orthogonal to the wave vector $\mathbf{k}$) decouple from the other three modes. Their evolution equation is
\begin{equation}
\label{6.7}
\frac{\partial \mathbf{w}_{\mathbf{k}\perp}}{\partial\tau}+\left(2\sqrt{2}\gamma_\text{s}^*+\frac{1}{2}\eta^*k^2\right)\mathbf{w}_{\mathbf{k}\perp}=0,
\end{equation}
where $\eta^*=\eta/\sigma^{1-d}\sqrt{mT_\text{s}}$. The solution to Eq.\ \eqref{6.7} is
\begin{equation}
\label{6.8}
\mathbf{w}_{\mathbf{k}\perp}(\mathbf{k},\tau)=\mathbf{w}_{\mathbf{k}\perp}(0)\exp\left[-\left(\frac{1}{2}\eta^*k^2+2\sqrt{2}\gamma_\text{s}^*
\right)\tau\right].
\end{equation}
Since both the (reduced) friction coefficient $\gamma_\text{s}^*$ and the (reduced) shear viscosity coefficient $\eta^*$ are positive, then the transversal shear modes of the granular suspension are linearly stable.

The remaining (longitudinal) modes correspond to $\rho_{\mathbf{k}}$, $\theta_{\mathbf{k}}$, and the longitudinal velocity component of the velocity field, $w_{\mathbf{k}\parallel}=\mathbf{w}_{\mathbf{k}}\cdot\widehat{\mathbf{k}}$ (parallel to $\mathbf{k}$). These modes are coupled and obey the equation
\begin{equation}
\label{6.9}
\frac{\partial\delta y_{\mathbf{k}\beta}(\tau)}{\partial\tau}+M_{\beta\mu}\delta y_{\mathbf{k}\mu}(\tau)=0,
\end{equation}
where $\delta y_{\mathbf{k}\beta}(\tau)$ denotes now the set $\left\{\rho_{\mathbf{k}}, w_{\mathbf{k}\parallel}, \theta_{\mathbf{k}}\right\}$ and $\mathsf{M}$ is the square matrix
\begin{equation}
\label{6.10}
\mathsf{M}=
\left(
\begin{array}{ccc}
0 & ik & 0  \\
ikp^*C_p & 2\sqrt{2}\gamma_\text{s}^*+\nu_\ell^*k^2 &ikp^*  \\
2\sqrt{2}\left(\zeta_0^*C_{\chi}+\zeta_0^{(1)}C_{n}+C_{\gamma}\right)+\mu^*k^2 &\frac{2}{d}ik\left(p^*+\frac{d}{2}\zeta_U\right) & 2\sqrt{2}\left(2\gamma_\text{s}^*\theta_\text{s}^{-1}+\frac{1}{2}\zeta_0^*+\zeta_0^{(1)}\theta_\text{s}\Delta_\theta\right)+D_{\text{T}}^*k^2 \\
\end{array}
\right).
\end{equation}
Here, the (reduced) transport coefficient $\nu_\ell^*$, $\mu^*$, and $D_{\text{T}}^*$ are defined as
\begin{equation}
\label{6.11}
\nu_\ell^*=\frac{1}{2\sigma^{1-d}\sqrt{mT_\text{s}}}\left(2\frac{d-1}{d}\eta+\eta_\text{b}\right), \quad
D_{\text{T}}^*=\frac{\kappa}{d\sigma^{1-d}\sqrt{T_\text{s}/m}}, \quad \mu^*=\frac{\rho}{d\sigma^{1-d}T_\text{s}\sqrt{mT_\text{s}}}\mu,
\end{equation}
while $p^*\equiv p_\text{s}/n T_\text{s}=1+2^{d-2}(1+\al)\chi \phi$, $\rho=m n$, and the quantities $C_p$, $C_\chi$, $C_n$, and $C_\gamma$ are given by
\begin{equation}
\label{6.12}
C_p=1+\phi\frac{\partial \ln p^*}{\partial\phi}, \quad C_{\chi}=1+\phi\frac{\partial \ln \chi}{\partial\phi},
\end{equation}
\begin{equation}
\label{6.13}
C_n=\phi\frac{\partial\chi}{\partial\phi} \Delta_\chi+\phi\frac{\partial\lambda}{\partial\phi}\Delta_\lambda, \quad
C_{\gamma}=2\left(1-\theta_\text{s}^{-1}\right)\gamma_\text{s}^*
\phi \frac{\partial \ln R}{\partial \phi}.
\end{equation}
In the above equations, it is understood that the transport coefficients $\eta^*$, $\nu_\ell^*$, $D_T^*$, and $\mu^*$ are evaluated in the homogeneous steady state.

The longitudinal three modes have the form $\text{exp}\left[\Lambda_{\ell}(k)\tau\right]$ for $\ell=1,2,3,$ where $\Lambda_{\ell}(k)$ are the eigenvalues of the matrix $\mathsf{M}$, namely, they are the solutions of the cubic equation
\begin{equation}
\label{6.14}
\Lambda^3+X(k)\Lambda^2+Y(k)\Lambda+Z(k)=0,
\end{equation}
where
\begin{equation}
\label{6.15}
X(k)=\sqrt{2}\left(\zeta_0^*+2\zeta_0^{(1)}\theta_\text{s}\Delta_\theta+4\gamma_\text{s}^*\theta_\text{s}^{-1}\right)
+k^2\left(D_{\text{T}}^*+\nu_\ell^*\right),
\end{equation}
\begin{equation}
\label{6.16}
Y(k)=\left(2\sqrt{2}\gamma_\text{s}^*+k^2\nu_\ell^*\right)\left[k^2 D_{\text{T}}^*+\sqrt{2}\left(\zeta_0^*+2\zeta_0^{(1)}\theta_\text{s} \Delta_\theta+4\gamma_\text{s}^*\theta_\text{s}^{-1}\right)\right]+k^2 p^*\left(C_p+\zeta_U+\frac{2}{d}p^*\right),
\end{equation}
\begin{equation}
\label{6.17}
Z(k)=p^*k^2\left[k^2\left(C_pD_{\text{T}}^*-\mu^*\right)+\sqrt{2}C_p\left(\zeta_0^*+2\zeta_0^{(1)}\theta_\text{s}\Delta_\theta+
4\gamma_\text{s}^*\theta_\text{s}^{-1}\right)-2\sqrt{2}\left(\zeta_0^*C_{\chi}+\zeta_0^{(1)}C_{n}+C_{\gamma}\right)\right].
\end{equation}
In general, one of the longitudinal modes can be unstable for $k<k_{\text{h}}$, where $k_{\text{h}}$ is obtained from Eq.\ \eqref{6.14} when $\Lambda=0$, namely, $Z(k_{\text{h}})=0$. The result is
\begin{equation}
\label{6.18}
k^2_{\text{h}}=\sqrt{2}\frac{2\left(\zeta_0^*C_{\chi}+\zeta_0^{(1)}C_{n}+C_{\gamma}\right)-C_p
\left(\zeta_0^*+2\zeta_0^{(1)}\theta_\text{s}\Delta_\theta+4\gamma_\text{s}^*\theta_\text{s}^{-1}\right)}{C_pD_{\text{T}}^*-\mu^*}.
\end{equation}
At a fixed value of the background temperature $T_\text{ex}^*$, a careful analysis of the dependence of $k_{\text{h}}^2$ on both the coefficient of restitution $\alpha$ and the volume fraction $\phi$ shows that $k_{\text{h}}^2$ is always negative. This means that there are no physical values of the wave numbers for which the longitudinal modes become unstable. Therefore, as in the case of the transversal shear modes, we can conclude that \emph{all} the eigenvalues of the dynamical matrix $\mathsf{M}$ have a \textit{positive} real part and no instabilities are found in the homogeneous steady state of a granular suspension.

\vicente{In summary, the stability analysis performed here by including the extra density dependencies of the transport coefficients shows no surprises relative to the earlier analysis \cite{GChV13}: the homogenous steady state of a moderately dense granular suspension is linearly stable. On the other hand, the dispersion relations derived here are different from those obtained in Ref.\ \cite{GChV13} since for instance the functional form of the heat flux transport coefficients differs in both approaches.}

\section{Conclusions}
\label{sec7}

In this paper we have undertaken a rather complete study on the transport properties of granular suspensions in the Navier--Stokes domain (first-order in the spatial gradients). The starting point of our study has been the Enskog kinetic equation where the effect of the gas phase on the solid particles is via the introduction of two additional terms: (i) a viscous drag force term proportional to the velocity of particle and (ii) a stochastic Langevin-like term. While the first term attempts to model the friction of solid particles on the viscous surrounding gas, the second term mimics the kinetic energy gained by grains due to eventual collisions with the more rapid molecules of the interstitial gas. Both terms are characterized by the friction coefficient $\gamma$ (which is a function of the volume fraction $\phi$) and the background temperature $T_\text{ex}$ (which is a known quantity of the model).

\vicente{A previous attempt on the derivation of the Navier--Stokes transport coefficients of dense granular suspensions was worked out by Garz\'o \emph{et al.} \cite{GChV13} by starting from a similar suspension model. However, the above work has two deficiencies: (i) it neglects an additional density dependence of the zeroth-order distribution $f^{(0)}$ through the parameter $\lambda(\phi)$ (defined in Eq.\ \eqref{3.9}), and (ii) it assumes that the friction coefficient $\gamma$ is constant. While the former simplification may be relevant in the evaluation of the diffusive heat conductivity coefficient (the transport coefficient associated to the density gradient in the heat flux), the latter simplification may be not reliable as the suspension becomes denser. The present analysis incorporates both extra new ingredients (the density dependence of $\lambda$ in $f^{(0)}$ and $\gamma=\gamma_0 R(\phi)$, $\gamma_0$ being constant) in the Chapman--Enskog solution. The results show that while these two new density dependencies do not formally affect the expression of the shear viscosity coefficient obtained in Ref.\ \cite{GChV13}, the forms of the heat flux transport coefficients and the cooling rate obtained here differ from those derived before. These findings are likely the most significant contributions of the present work. In this context, this paper complements and extends previous papers on transport properties in granular suspensions \cite{GTSH12,GChV13,GFHY16}.}

Before considering inhomogeneous situations, the homogeneous steady state has been analyzed. As expected, after a transient period, the steady distribution function $f_\text{s}$ adopts the form \eqref{3.8} where the temperature dependence of the scaled distribution $\varphi_\text{s}$ is encoded through the dimensionless velocity $\mathbf{c}=\mathbf{v}/v_0$ ($v_0=\sqrt{2T_\text{s}/m}$ being the thermal speed) and the (scaled) friction coefficient $\gamma_\text{s}^*=\lambda (\phi) \theta_\text{s}^{-1/2}$ ($\theta_\text{s}=T_\text{s}/T_\text{ex}$ being the reduced steady temperature). As in previous works on granular fluids driven by thermostats \cite{GMT12,GChV13}, the above scaling differs from the one assumed for undriven granular fluids \cite{NE98,BP04,G19} where $\varphi_\text{s}$ depends on $T$ only through the scaled velocity $\mathbf{c}$. Although the exact form of $\varphi_\text{s}$ is not known, a good approximation of this distribution (at least in the thermal velocity region $c \sim 1$) is provided by the leading Sonine approximation \eqref{3.12}. By using this distribution, we have explicitly obtained the fourth cumulant $a_{2,\text{s}}$; this coefficient provides an indirect information on the deviation of $\varphi_\text{s}$ from its Maxwellian form $\pi^{-d/2} e^{-c^2}$. Once $a_{2,\text{s}}$ is known, the steady temperature $\theta_\text{s}$ is obtained by solving the cubic equation \eqref{3.16}. In spite of the above approximations, the theoretical predictions for $\theta_\text{s}$ and $a_{2,\text{s}}$ show an excellent agreement with Monte Carlo simulation results. \vicente{As expected, the results obtained for homogeneous systems agree with those derived in Ref.\ \cite{GChV13} when one makes the mapping $\xi_\text{s}^*\to 2\lambda \theta_\text{s}^{-3/2}$ with $R(\phi)=1$.}

Once the steady reference state is well characterized, we have considered the transport processes occurring in granular suspensions with small spatial gradients of the hydrodynamic fields. In this situation, the Enskog kinetic equation has been solved by means of the Chapman--Enskog method \cite{CC70} where only terms up to the first order in the spatial gradients have been retained (Navier--Stokes hydrodynamic order). As in previous papers on the application of the Chapman--Enskog method to granular systems \cite{BDKS98,GD99a,L05,GChV13}, the spatial gradients have been assumed to be independent of the coefficient of restitution $\al$. Thus, although the constitutive equations for the irreversible fluxes are limited to first order in spatial gradients, the corresponding transport coefficients appearing in these equations apply \emph{a priori} to arbitrary degree of collisional dissipation. This type of expansion differs from the ones considered by other authors \cite{GS96,SGN96,SG98,GNB05} where the Chapman--Enskog solution is given in powers of both the Knudsen number (or spatial gradients as in the conventional scheme) and the degree of collisional dissipation $\delta \equiv 1-\al^2$. The results reported here are consistent with the ones obtained in those papers \cite{GS96,SGN96,SG98,GNB05} in the limit $\delta\to 0$.

\vicente{As in the Chapman--Enskog solution obtained in Ref.\ \cite{GChV13}}, a subtle but important point is the choice of the zeroth-order approximation $f^{(0)}$ in the perturbation expansion. Although we are interested in obtaining the transport coefficients in steady state conditions, for general small perturbations around the homogeneous steady state, the density and temperature are specified separately in the local reference state $f^{(0)}$ and hence, it is not expected that the temperature is stationary at any point of the system. This means that $\partial_t^{(0)}T \neq 0$ in the reference base state and consequently, the complete determination of the Navier--Stokes transport coefficients requires to know for instance the temperature dependence of the fourth cumulant $a_2$ of the \emph{unsteady} reference state. This of course involves the numerical integration of the differential equation \eqref{4.13.1}. This is quite an intricate problem that goes beyond the objective of this paper. Since we are essentially motivated by a desire for analytic expressions, the steady state conditions have been considered. On the other hand, given that $\partial_t^{(0)}T \neq 0$ in the Chapman--Enskog scheme, the transport coefficients are defined not only in terms of the hydrodynamic fields in the steady state but also there are contributions to the transport coefficients [such as the derivatives $\Delta_\theta$, $\Delta_\lambda$, and $\Delta_\chi$ defined by Eqs.\ \eqref{4.14}--\eqref{4.16}, respectively] accounting for the vicinity of the perturbed state to the steady state.

As usual, in order to obtain explicit expressions for the transport coefficients, the leading terms in a Sonine polynomial expansion have been considered. These forms have been displayed along the section \ref{sec5}: the bulk $\eta_\text{b}$ and shear $\eta$ viscosities are given by Eqs.\ \eqref{5.3} and \eqref{5.5}, respectively, the thermal conductivity $\kappa$ is given by Eqs.\ \eqref{5.6} and \eqref{5.8}, the heat diffusive conductivity $\mu$ is given by Eqs.\ \eqref{5.10} and \eqref{5.11} and the first-order contribution $\zeta_U$ to the cooling rate is given by Eqs.\ \eqref{5.14} and \eqref{5.15}. \vicente{As said before, the expressions of $\eta_\text{b}$ and $\eta$ agree with those derived in \cite{GChV13} (once one takes $R(\phi)=1$) while the expressions of $\kappa$, $\mu$, and $\zeta_U$ reduce to those obtained in \cite{GChV13} when the contributions coming from the derivatives $\Delta_\theta$, $\Delta_\lambda$, and $\Delta_\chi$ are neglected.}

In reduced forms, it is quite apparent that the Navier--Stokes coefficients of the granular suspension exhibit a complex dependence on the (steady) temperature $\theta_\text{s}$, the coefficient of restitution $\al$, the solid volume fraction $\phi$, and the (reduced) background temperature $T_\text{ex}^*$. In addition, Figs.\ \eqref{fig6}--\eqref{fig8} highlight the significant impact of the gas phase on the Navier--Stokes transport coefficients $\eta$, $\kappa$, and $\mu$ since their $\al$-dependence is clearly different from the one previously found for dry granular gases \cite{BDKS98,GD99a}.

As an application of the previous results, the stability of the special homogeneous steady state solution has been analyzed. This has been
achieved by solving the linearized Navier--Stokes hydrodynamic equations for small perturbations around the homogeneous steady state. The
linear stability analysis performed here shows no new surprises relative to the earlier work \cite{GChV13}: the homogeneous steady state is linearly stable with respect to long enough wavelength excitations \vicente{(namely, long enough small spatial gradients)}. On the other hand, it is worthwhile to recall that the  conclusion reached here for the reference homogeneous steady state differs from the one found for freely cooling granular fluids where it was shown \cite{BDKS98,G05} that the resulting hydrodynamic equations exhibit a long wavelength instability for three of the hydrodynamic modes. This shows again the influence of the interstitial fluid on the dynamics of solid particles.

It is quite apparent that the theoretical results obtained in this paper under certain approximations should be tested
against computer simulations. This would allow us to gauge the degree of accuracy of the theoretical predictions. As
happens for dry granular gases \cite{BRM98,BRC99,BR04,BRMG05,MSG05,MSG07,MDCPH11,MGHEH12,BR13,MGH14}, we expect that
the present results stimulate the performance of appropriate simulations where the kinetic theory calculations reported here can be assessed. We also plan to undertake such kind of simulations for the case of the shear viscosity. More specifically, we want to perform simulations of granular suspensions under uniform shear flow where the Navier--Stokes shear viscosity might be measured in the Newtonian regime (very small shear rates). Another possible project for the next future is the extension of the present results to the relevant subject of multicomponent granular suspensions. Work along these lines will be worked out in the near future.

\acknowledgments

We want to thank Mois\'es Garc\'ia Chamorro for providing us the simulation data included in Figs.\ \ref{fig1}--\ref{fig4}. The present work has been supported by the Spanish Government through Grant No. FIS2016-76359-P and by the Junta de Extremadura (Spain)
Grant Nos. IB16013 (V.G.) and GR18079, partially financed by ``Fondo Europeo de Desarrollo Regional'' funds. The research of Rub\'en G\'omez Gonz\'alez has been supported by the predoctoral fellowship BES-2017-079725 from the Spanish Government.

\appendix
\section{Some technical details on the first-order solution}
\label{appA}

Up to the first order in the expansion, the velocity distribution function $f^{(1)}$ verifies the Enskog kinetic equation
\begin{equation}
\label{a1}
\partial_t^{(0)}f^{(1)}-\gamma\frac{\partial}{\partial\mathbf{v}}\cdot\mathbf{V}f^{(1)}-\frac{\gamma T_{\text{ex}}}{m}\frac{\partial^2 f^{(1)}}{\partial v^2}=-\left(D_t^{(1)}+\mathbf{V}\cdot\nabla\right)f^{(0)}+\gamma \Delta \mathbf{U}\cdot \frac{\partial f^{(0)}}{\partial \mathbf{v}}+J_{\text{E}}^{(1)}[f,f],
\end{equation}
where $D_t^{(1)}\equiv \partial_t^{(1)}+\mathbf{U}\cdot \nabla$ and $J_{\text{E}}^{(1)}[f,f]$ denotes the first-order contribution to the expansion of the Enskog collision operator in powers of the spatial gradients. In order to explicitly determine $J_{\text{E}}^{(1)}[f,f]$ we need the results
\begin{equation}
\label{a2}
\chi\left(\mathbf{r},\mathbf{r}\pm\boldsymbol{\sigma}|n\right)\rightarrow\chi\left(1\pm\frac{1}{2}n\frac{\partial \ln \chi}{\partial n}\boldsymbol{\sigma}\cdot\nabla\ln n\right),
\end{equation}
\begin{equation}
\label{a3}
f^{(0)}(\mathbf{r}\pm\boldsymbol{\sigma},\mathbf{v}; t)\rightarrow f^{(0)}(\mathbf{r},\mathbf{v}; t)\pm f^{(0)}(\mathbf{r},\mathbf{v}; t)
\left[n \frac{\partial f^{(0)}}{\partial n}\boldsymbol{\sigma}\cdot\nabla\ln n +T \frac{\partial f^{(0)}}{\partial T}\boldsymbol{\sigma}\cdot\nabla\ln T-\frac{\partial f^{(0)}}{\partial V_i}(\boldsymbol{\sigma}\cdot\nabla)U_i\right],
\end{equation}
where $\chi$ is obtained from the functional $\chi(\mathbf{r},\mathbf{r}\pm\boldsymbol{\sigma}|n)$ by evaluating all density fields at $n(\mathbf{r}, t)$. Taking into account Eqs.\ \eqref{a2} and \eqref{a3}, $J_{\text{E}}^{(1)}$ reads \cite{GChV13}
\beqa
\label{a4}
J_{\text{E}}^{(1)}[f,f]&=& -\boldsymbol{\mathcal{K}}\Big[n\frac{\partial f^{(0)}}{\partial n}\Big]\cdot\nabla\ln n-\frac{1}{2}\phi\left(\frac{\partial\ln\chi}{\partial \phi}\right)\boldsymbol{\mathcal{K}}\Big[f^{(0)}\Big]\cdot\nabla\ln n
-\boldsymbol{\mathcal{K}}\Big[T\frac{\partial f^{(0)}}{\partial T}\Big]\cdot\nabla\ln T
\nonumber\\
& & +\frac{1}{2}\mathcal{K}_i\left[\frac{\partial f^{(0)}}{\partial V_j}\right]\left(\frac{\partial U_i}{\partial r_j}+\frac{\partial U_j}{\partial r_i}-\frac{2}{d}\delta_{ij}\nabla\cdot\mathbf{U}\right)+\frac{1}{d}\mathcal{K}_i\left[\frac{\partial f^{(0)}}{\partial V_i}\right]\nabla\cdot\mathbf{U}-\mathcal{L}f^{(1)},
\eeqa
where $\mathcal{L}$ is defined by Eq.\ \eqref{4.21} and the operator $\boldsymbol{\mathcal{K}}[X]$ is given by
\begin{equation}
\label{a5}
\boldsymbol{\mathcal{K}}[X]=\sigma^d\chi\int d \mathbf{v}_2\int d\widehat{\boldsymbol{\sigma}}\Theta(\widehat{\boldsymbol{\sigma}}
\cdot\mathbf{g}_{12})\left(\widehat{\boldsymbol{\sigma}}\cdot\mathbf{g}_{12}\right)\widehat{\boldsymbol{\sigma}}
\left[\alpha^{-2}f^{(0)}(\mathbf{v}_1'')X(\mathbf{v}_2'')+f^{(0)}(\mathbf{v}_1)X(\mathbf{v}_2)\right].
\end{equation}
As already noted in Ref. \cite{GChV13}, upon obtaining Eq.\ \eqref{a4} use has been made of the symmetry property $\mathcal{K}_i[\partial_{V_j}
f^{(0)}] =\mathcal{K}_j[\partial_{V_i}f^{(0)}]$ that follows from the isotropy of the zeroth-order solution. Thus we are
able to separate the contributions from the flow field gradients into independent traceless and diagonal components.

The macroscopic balance equations to first order in the gradients are
\begin{equation}
\label{a6}
D_t^{(1)}n=-n\nabla\cdot\mathbf{U},\quad D_t^{(1)}\mathbf{U}=-\rho^{-1}\nabla p-\gamma \Delta \mathbf{U}, \quad
D_t^{(1)}T=-\frac{2p}{dn}\nabla\cdot\mathbf{U}-\zeta^{(1)}T,
\end{equation}
where $\zeta^{(1)}$ is the first order contribution to the cooling rate. Since the cooling rate is a scalar, corrections to first-order in the gradients can arise only from $\nabla \cdot \mathbf{U}$ since $\nabla n$ and $\nabla T$ are vectors and the tensor $\partial_j U_i+\partial_i U_j-\frac{2}{d}\delta_{ij}\nabla \cdot \mathbf{U}$ is a traceless tensor. Thus, $\zeta^{(1)}$ can be written as
\beq
\label{a12.1}
\zeta^{(1)}=\zeta_U \nabla \cdot \mathbf{U}.
\eeq
\vicente{The unknown quantity $\zeta_U$ is a functional of the first-order distribution $f^{(1)}$. A more explicit form for $\zeta_U$ is obtained by expanding Eq.\ \eqref{2.21} to first-order in gradients. This yields Eq.\ \eqref{5.13} where $\zeta_1^{(0)}$ and $\zeta_1^{(1)}$ are defined by Eqs.\ \eqref{5.14} and \eqref{b16}, respectively.}

The use of the balance equations \eqref{a6} allows us to evaluate the right-hand side of Eq.\ \eqref{a1}. The combination of these results with the forms \eqref{a4} of the Enskog collision operator $J_{\text{E}}^{(1)}$ \vicente{and \eqref{5.13} of $\zeta_U$} leads to the expression
\beqa
\label{a7}
\left(\partial_t^{(0)}+\mathcal{L}\right)f^{(1)}-\gamma\frac{\partial}{\partial\mathbf{v}}\cdot\mathbf{V}f^{(1)}-\frac{\gamma T_{\text{ex}}}{m}\frac{\partial^2 f^{(1)}}{\partial v^2}&-& \zeta_1^{(1)}T\frac{\partial f^{(0)}}{\partial T}\nabla \cdot \mathbf{U}=\mathbf{A}\cdot\nabla\ln T+\mathbf{B}\cdot\nabla\ln n \nonumber\\
& & +C_{ij}\frac{1}{2}\left(\frac{\partial U_i}{\partial r_j}+\frac{\partial U_j}{\partial r_i}-\frac{2}{d}\delta_{ij}\nabla\cdot\mathbf{U}\right)+D\nabla\cdot\mathbf{U},
\eeqa
where
\beq
\label{a8}
\mathbf{A}(\mathbf{V})=-\mathbf{V}T\frac{\partial f^{(0)}}{\partial T}-\frac{p}{\rho}\frac{\partial f^{(0)}}{\partial\mathbf{V}}-\boldsymbol{\mathcal{K}}\Big[T\frac{\partial f^{(0)}}{\partial T}\Big],
\eeq
\beq
\label{a9}
\mathbf{B}(\mathbf{V})=-\mathbf{V}n\frac{\partial f^{(0)}}{\partial n}-\frac{p}{\rho}\left(1+\phi\frac{\partial \ln p^*}{\partial\phi}\right)\frac{\partial f^{(0)}}{\partial\mathbf{V}}-\boldsymbol{\mathcal{K}}\Big[n
\frac{\partial f^{(0)}}{\partial n}\Big]-\frac{1}{2}\phi\left(\frac{\partial \ln\chi}{\partial\phi}\right)\boldsymbol{\mathcal{K}}\Big[f^{(0)}\Big],
\eeq
\beq
\label{a10}
C_{ij}(\mathbf{V})=V_i\frac{\partial f^{(0)}}{\partial V_j}+\mathcal{K}_i\Big[\frac{\partial f^{(0)}}{\partial V_j}\Big],
\eeq
\beq
\label{a11}
D(\mathbf{V})=\frac{1}{d}\frac{\partial}{\partial\mathbf{V}}\cdot\left(\mathbf{V}f^{(0)}\right)+\left(\zeta_1^{(0)}+\frac{2}{d}p^*\right)
T\frac{\partial f^{(0)}}{\partial T}-f^{(0)}+n\frac{\partial f^{(0)}}{\partial n}+\frac{1}{d}\mathcal{K}_i\Big[\frac{\partial f^{(0)}}{\partial V_i}\Big].
\eeq
Here, $p^*\equiv p/(n T)$. The structure of Eqs.\ \eqref{a7}--\eqref{a11} is formally equivalent to the ones derived for driven granular gases \cite{GChV13}. The only difference lies on the dependence of the zeroth-order solution $f^{(0)}$ on density and temperature.

As for dry granular gases \cite{GD99a}, the solution to the kinetic equation \eqref{a7} is given by Eq.\ \eqref{4.16.1} where the unknown functions $\boldsymbol{\mathcal{A}}$, $\boldsymbol{\mathcal{B}}$, $\mathcal{C}_{ij}$, and $\mathcal{D}$ are determined by solving Eq.\ \eqref{a7}. Since the gradients of the hydrodynamic fields are all independent, substitution of \eqref{4.16.1} into Eq.\ \eqref{a7} yields a set of linear, inhomogeneous integral equations. In order to obtain them, one needs the result
\beqa
\label{a12}
\partial_t^{(0)}\nabla\ln T&=&\nabla\partial_t^{(0)}\ln T=\nabla\bigg(2\gamma\left(\theta^{-1}-1\right)-\zeta^{(0)}\bigg)=-\Bigg[\zeta^{(0)}\left(1+\phi\frac{\partial\ln\chi}{\partial\phi}\right)
+\chi\phi\frac{\partial\chi}{\partial\phi}\frac{\partial}{\partial\chi}\left(\frac{\zeta^{(0)}}{\chi}\right)
\nonumber\\
& &
-\lambda\left(1-\phi\frac{\partial\ln R}{\partial \phi}\right)\frac{\partial \zeta^{(0)}}{\partial \lambda}-2\left(\theta^{-1}-1\right)\gamma \phi\frac{\partial\ln R}{\partial \phi}\Bigg]\nabla\ln n\nonumber\\
& & -\left(2\gamma\theta^{-1}+\frac{1}{2}\zeta^{(0)}+\zeta^{(0)}\theta \frac{\partial\ln \zeta^*_0}{\partial\theta}
\right)\nabla\ln T.
\eeqa
The integral equations \eqref{4.17}--\eqref{4.20} can be easily obtained after taking into account Eq.\ \eqref{a12} and the steady state condition $\Lambda^{(0)}=0$.

\section{Kinetic contributions to the transport coefficients}
\label{appB}

In this Appendix we give some details on the determination of the kinetic contributions to the transport coefficients $\eta$, $\kappa$, and $\mu$ as well as the first-order contribution $\zeta_U$ to the cooling rate. Since all these quantities are obtained int he steady state, the subscript $\text{s}$ appearing along the main text will be omitted here for the sake of brevity.

The kinetic part of the shear viscosity $\eta_\text{k}$ is defined as
\beq
\label{b1}
\eta_\text{k}=-\frac{1}{(d-1)(d+2)}\int d \mathbf{v}\; D_{ij}(\mathbf{V})\;C_{ij}(\mathbf{V}),
\eeq
where
\begin{equation}
\label{b2}
D_{ij}=m\left(V_iV_j-\frac{1}{d}V^2\delta_{ij}\right).
\end{equation}
As usual, to get $\eta_\text{k}$ one multiplies both sides of Eq.\ \eqref{4.17} by $D_{ij}$ and integrates over velocity. The result is
\beq
\label{b3}
\left(2\gamma+\nu_{\eta}\right)\eta_\text{k}=nT-\frac{1}{(d-1)(d+2)}\int d\mathbf{V}D_{ij}(\mathbf{V})
\mathcal{K}_i\left[\frac{\partial f^{(0)}}{\partial V_j}\right],
\eeq
where
\beq
\label{b4}
\nu_{\eta}=\frac{\int d\mathbf{v} D_{ij}(\mathbf{V})\mathcal{L}\mathcal{C}_{ij}(\mathbf{V})}{\int d\mathbf{v} D_{ij}(\mathbf{V})\mathcal{C}_{ij}(\mathbf{V})},
\eeq
and \cite{GD99a,L05,G13}
\beq
\label{b5}
\int d \mathbf{V} D_{ij}(\mathbf{V})\mathcal{K}_i\left[\frac{\partial f^{(0)}}{\partial V_j}\right]=2^{d-2}(d-1)nT\chi\phi(1+\alpha)(1-3\alpha).
\eeq
The expression of $\eta_\text{k}$ can be easily obtained when one takes into account Eq.\ \eqref{b5} and the explicit form \eqref{5.5.1} of $\nu_\eta$. This latter expression is obtained from Eq.\ \eqref{b4} by considering the leading terms in a Sonine polynomial expansion of the unknown $\mathcal{C}_{ij}(\mathbf{V})$.

The kinetic parts $\kappa_\text{k}$ and $\mu_\text{k}$ are defined, respectively, as
\begin{equation}
\label{b6}
\kappa_\text{k}=-\frac{1}{dT}\int d\mathbf{v} \mathbf{S}(\mathbf{V})\cdot\boldsymbol{\mathcal{A}}(\mathbf{V}),
\end{equation}
\begin{equation}
\label{b7}
\mu_\text{k}=-\frac{1}{dn}\int d\mathbf{v}\mathbf{S}(\mathbf{V})\cdot\boldsymbol{\mathcal{B}}(\mathbf{V}),
\end{equation}
where
\begin{equation}
\label{b8}
\mathbf{S}(\mathbf{V})=\left(\frac{m}{2}V^2-\frac{d+2}{2}T\right)\mathbf{V}.
\end{equation}
As in the case of $\eta_\text{k}$, $\kappa_\text{k}$ is obtained by multiplying both sides of Eq.\ \eqref{4.17} by $\mathbf{S}(\mathbf{V})$ and integrating over $\mathbf{v}$. The result is
\beq
\label{b9}
\left(\nu_\kappa +\gamma\theta^{-1}-2\zeta^{(0)}-\zeta^{(0)}\theta \frac{\partial\ln \zeta^*_0}{\partial\theta}
\right)\kappa_\text{k}=-\frac{1}{d T}\int d\mathbf{V}\mathbf{S}(\mathbf{V})\cdot\mathbf{A},
\eeq
where use has been made of the steady state condition \eqref{3.7} and
\beq
\label{b10}
\nu_{\kappa}=\frac{\int d\mathbf{v}\mathbf{S}(\mathbf{V})\cdot\mathcal{L}\boldsymbol{\mathcal{A}}(\mathbf{V})}
{\int d\mathbf{v}\mathbf{S}(\mathbf{V})\cdot\boldsymbol{\mathcal{A}}(\mathbf{V})}.
\eeq
The right hand side of Eq.\ \eqref{b9} can be computed when one takes into account Eq.\ \eqref{a8} and the relationship \eqref{4.9}. After some algebra, one gets
\beqa
\label{b11}
-\frac{1}{dT}\int d\mathbf{V}\mathbf{S}\cdot\mathbf{A}&=&\frac{1}{dT}\bigg\{\frac{d(d+2)}{2m}nT^2\left(1+2a_2+\theta
\Delta_\theta\right)-\frac{1}{2}\int d\mathbf{v}\mathbf{S}(\mathbf{V})\cdot\boldsymbol{\mathcal{K}}\left[\frac{\partial}
{\partial\mathbf{V}}\cdot\left(\mathbf{V}f^{(0)}\right)\right]\nonumber\\
& & +\frac{\theta}{a_2}\Delta_\theta \int d\mathbf{v}\mathbf{S}(\mathbf{V})\cdot\boldsymbol{\mathcal{K}}\left[f^{(0)}-f_{\text{M}}\right]\bigg\},
\eeqa
where $f_{\text{M}}(\mathbf{c})=n\pi^{-d/2}v_0^{-d}\text{e}^{-c^2}$ and use has been made of the Sonine approximation \eqref{3.12} to $f^{(0)}$ and the property \eqref{4.9}. The first collision integral involving the operator $\boldsymbol{\mathcal{K}}$ has been calculated in previous works \cite{GD99a,L05,G13} and the result is
\beq
\label{b12}
\int d\mathbf{V}\mathbf{S}(\mathbf{V})\cdot\boldsymbol{\mathcal{K}}\left[\frac{\partial}
{\partial\mathbf{V}}\cdot\left(\mathbf{V}f^{(0)}\right)\right]=-\frac{3}{8}2^d d\frac{nT^2}{m}\chi\phi\left(1+\alpha\right)^2
\left[2\alpha-1+a_2(1+\alpha)\right].
\eeq
The second collision integral in \eqref{b11} has not been evaluated before. After some algebra, one gets
\beq
\label{b13}
\int d\mathbf{V}\mathbf{S}(\mathbf{V})\cdot\boldsymbol{\mathcal{K}}\left[f^{(0)}-f_{\text{M}}\right]=
\frac{3}{32}2^d d\frac{nT^2}{m}\chi\phi\left(1+\alpha\right)^3a_2.
\eeq
With the above results, $\kappa_\text{k}$ can be finally written in the form \eqref{5.8}. As in the case of $\nu_\eta$, the (reduced) collision frequency $\nu_\kappa$ can be well estimated by considering the leading Sonine approximation to $\boldsymbol{\mathcal{A}}$.

The evaluation of $\mu_\text{k}$ follows similar mathematical steps to those made for $\kappa_\text{k}$ since one has to multiply both sides of Eq.\ \eqref{4.18} by $\mathbf{S}(\mathbf{V})$ and integrate over $\mathbf{v}$. In order to get its explicit form \eqref{5.11}, one needs the partial results
\beqa
\label{b14}
-\frac{1}{dn}\int d\mathbf{V}\mathbf{S}\cdot\mathbf{B}&=&\frac{d+2}{2}\frac{T^2}{m}\left[a_2-\lambda\left(1-\phi \frac{\partial \ln R}{\partial \phi}\right)\Delta_\lambda+\phi\frac{\partial\chi}{\partial\phi}\Delta_\chi\right]\nonumber\\
& & -\frac{a_2^{-1}}{d n}\left[\lambda\left(1-\phi \frac{\partial \ln R}{\partial \phi}\right)\Delta_\lambda-\phi\frac{\partial\chi}{\partial\phi}\Delta_\chi\right]
\int d\mathbf{V}\mathbf{S}(\mathbf{V})\cdot\boldsymbol{\mathcal{K}}\left[f^{(0)}-f_{\text{M}}\right]
\nonumber\\
& & +\frac{1}{dn}\left(1+\frac{1}{2}\phi\frac{\partial \ln\chi}{\partial \phi}\right)\int d\mathbf{V}\mathbf{S}(\mathbf{V})\cdot\boldsymbol{\mathcal{K}}\left[f^{(0)}\right],
\eeqa
\beq
\label{b15}
\int d\mathbf{V}\mathbf{S}(\mathbf{V})\cdot\boldsymbol{\mathcal{K}}\left[f^{(0)}\right]=
\frac{3}{8}2^d d\frac{nT^2}{m}\chi\phi\left(1+\alpha\right)\left[\alpha\left(\alpha-1\right)+
\frac{a_2}{6}\left(10+2d-3\alpha+3\alpha^2\right)\right].
\eeq
The expression \eqref{5.11} can be derived by using Eqs.\ \eqref{b14} and \eqref{b15}.

Finally, the contribution $\zeta_1^{(1)}$ to the cooling rate $\zeta_U$ is defined as
\beq
\label{b16}
\zeta_1^{(1)}=\frac{1}{2nT}\frac{\pi^{(d-1)/2}}{d\Gamma\left(\frac{d+3}{2}\right)}\sigma^{d-1}\chi m\left(1-\alpha^2\right)\int d\mathbf{V}_1\int d\mathbf{V}_2 g^3f^{(0)}(\mathbf{V}_1)\mathcal{D}(\mathbf{V}_2),
\eeq
where the unknown function $\mathcal{D}(\mathbf{V})$ is the solution of the linear integral equation \eqref{4.20}. As before, an approximate solution to \eqref{4.20} can be obtained by taking the Sonine approximation
\begin{equation}
\label{b17}
\mathcal{D}(\mathbf{V})\rightarrow e_D\; f_{\text{M}}(\mathbf{V})\; F(\mathbf{V}),
\end{equation}	
where
\begin{equation}
\label{b18}
F(\mathbf{V})=\left(\frac{m}{2T}\right)^2V^4-\frac{d+2}{2}\frac{m}{T}V^2+\frac{d\left(d+2\right)}{4}.
\end{equation}
The coefficient $e_D$ is given by
\begin{equation}
\label{b19}
e_D=\frac{2}{d\left(d+2\right)}\frac{1}{n}\int d\mathbf{V}\mathcal{D}(\mathbf{V})F(\mathbf{V}).
\end{equation}
Substitution of Eq.\ \eqref{b17} into Eq.\ \eqref{b16} gives
\begin{equation}
\zeta_1^{(1)}=\frac{3\left(d+2\right)}{32d}\chi\left(1-\alpha^2\right)\left(1+\frac{3}{32}a_2\right)\nu_0e_D,
\end{equation}
where $\nu_0=nT/\eta_0$. The coefficient $e_D$ is obtained by substituting the Sonine solution \eqref{b17} into the integral equation \eqref{4.20}, multiplying it by the polynomial $F(\mathbf{V})$ and integrating over velocity. After some algebra one gets the expression \eqref{5.15} for $\zeta_1^{(1)}$.

%\bibliography{suspension_NS}
%\end{document}

%\bibliography{suspension_NS}
\end{document}